\documentclass[11pt,a4paper]{article}
\usepackage{epsfig}

\newlength{\dinwidth}
\newlength{\dinmargin}
\def\eq#1{{(\ref{#1})}}

\newcommand{\beq}{\begin{equation}}
\newcommand{\eeq}{\end{equation}}
\newcommand{\beqar}[1]{\begin{eqnarray}\label{#1}}
\newcommand{\eeqar}{\end{eqnarray}}

\def\fig#1{{Fig.~\ref{#1}}}
\newcommand{\tdm}[1]{\mbox{\boldmath $#1$}}

\newcommand{\vb}{\tdm{b}}

\newcommand{\fda}{f^\dagger}
\newcommand{\kbf}{{\cal K}_0}
\newcommand{\fbk}{f^{\mathrm{BK}}(y,k^2)}
\newcommand{\fbkd}{f^{\mathrm{\dagger BK}}(y',k^2)}
\newcommand{\beeq}{\begin{eqnarray}}
\newcommand{\eeeq}{\end{eqnarray}}
\newcommand{\be}{\begin{equation}}
\newcommand{\ee}{\end{equation}}

\def\kbo{{\bf k}}

\def\rbo{{\bf r}}
\def\rbop{{\bf r}^\prime}
\def\bbo{{\bf b}}

\def\ytau{y}

\begin{document}

\title{{~}\\[1cm]
{\Large \bf 
Solving effective field theory of interacting QCD pomerons 
\\ in the semi-classical approximation }}
\author{ 
{~}\\
S.~Bondarenko$\,{}^{a)}\,$\thanks{Email: sergb@mail.desy.de}, 
\hspace{1ex}
L.~Motyka$\,{}^{b),c)}\,$\thanks{E-mail: motyka@th.if.uj.edu.pl}
\\[10mm]
{\it\normalsize ${}^{a)}$ II Institute for  Theoretical Physics, 
University of Hamburg, Germany}\\
{\it\normalsize ${}^{b)}$ DESY Theory Group, Hamburg, Germany }\\
{\it\normalsize $^{c)}$ Institute of Physics, Jagellonian University,
Krak\'{o}w, Poland} \\}

\date{16 May 2006}

\maketitle
\thispagestyle{empty}

\begin{abstract}
Effective field theory of BFKL pomerons interacting by QCD 
triple pomeron vertices is investigated. Classical equations of motion 
for the effective pomeron fields are presented being a minimal extension 
of the Balitsky-Kovchegov equation that incorporates both  
merging and splitting of the pomerons and that is self-dual.   
The equations are solved for symmetric boundary conditions. 
The solutions provide the dominant contribution to the scattering 
amplitudes in the semi-classical approximation. We find that for 
rapidities of the scattering larger than a critical value $Y_c$ at 
least two classical solutions exist. Curiously, for each of the two 
classical solutions with the lowest action the symmetry between 
the projectile and the target is found to be spontaneously broken, 
being however preserved for the complete set of classical solutions. 
The solving configurations at rapidities $Y>Y_c$ consist of a Gribov 
field being strongly suppressed even at very large gluon momenta and the 
complementary Gribov field that converges at high $Y$ to a solution of 
Balitsky-Kovchegov equation. Interpretation of the results is given 
and possible consequences are shortly discussed. 
\end{abstract}

\begin{flushright}
\vspace{-20.8cm}
%{DESY 06--xxx}
hep-ph/0605185\\
%\today
\end{flushright}
\thispagestyle{empty}

\newpage

\section{Introduction}

Understanding of scattering of hadrons and nuclei at very high energies 
in terms of Quantum Chromodynamics remains one of the most important
challenges for the theory of strong interactions. The reasons for that are
both of the phenomenological and of the purely theoretical nature. 
In practical terms, the basic physics of the present and future 
colliders, HERA, Tevatron, RHIC and the CERN Large Hadron Collider (LHC), 
is the physics of high energy scattering in QCD. 
On the other hand, the high energy regime of QCD reveals intriguing 
similarities to high energy regime of string theory~\cite{ads1}. 
To trace and explore postulated dualities between string theories and 
gauge theories and possible manifestations of those dualities in high 
energy scattering is a goal of primary importance.

The most successful approach to high energy scattering in QCD is based
on infinite resummations of large logarithms of collision energy $\sqrt{s}$
in perturbative expansions of the scattering amplitudes. The corner stone 
of this formalism is the evolution equation derived by Balitsky, Fadin, 
Kuraev and Lipatov (BFKL)\cite{bfkl,bfklsum,nlbfkl}. 
In the BFKL framework, the energy evolution of the exchange of an 
interacting gluon pair in a colour singlet state (the BFKL pomeron) 
was analyzed at the leading logarithmic (LL) approximation, and consequently, 
the energy dependence of the hard scattering amplitude was obtained that
exhibits a power like growth with energy, ${\cal A} \sim s^{1+\Delta}$ 
with the intercept $\Delta \sim 0.3$. Such behaviour would eventually lead 
to violation of unitarity. Clearly, this indicates that when the energy 
is sufficiently large unitarity corrections to the BFKL evolution 
must be added.

In recent years unitarity effects in high energy QCD have been vigorously 
investigated along two main lines. The Color Glass Condensate (CGC) approach
\cite{balitsky,jimwalk} is formulated in the transverse position space and it 
is based on energy evolution of Wilson loops and phenomena of rescattering 
and recombination. In the large $N_c$ limit, the dynamics of the 
Color Glass Condensate may be analyzed in terms of a statistical model 
of color dipoles~\cite{dipmod}.  At a very general level, the dipole 
description of high energy scattering may be presented as a combination of 
multiple dipole splittings, the rescattering of dipoles off a target and 
a stochastic fluctuation term~\cite{stochastic}.

The QCD Reggeon Field Theory (QCD-RFT) approach is formulated in momentum 
space and bases the on standard diagrammatic 
calculus~\cite{vert1,vert2,eglla1,eglla2,eglla3,eglla4}. The main building 
blocks here are QCD reggeon Green's functions and multi-reggeon vertices 
derived in the perturbative QCD. Scattering amplitudes may be represented 
in terms of Feynman diagrams of effective (non-local) reggeon fields. 
This effective field theory is constructed using the so-called 
Extended Generalized Leading Logarithmic Approximation (EGLLA)~\cite{eglla1} 
which resums the leading powers of $\,\alpha_s \log s\,$ for given topology of 
the reggeon diagram. 
The CGC and the QCD-RFT formulations should be, in fact, 
two different descriptions of the same theory, and they 
should be equivalent.

\begin{figure}[t]
\begin{center}
\psfig{file=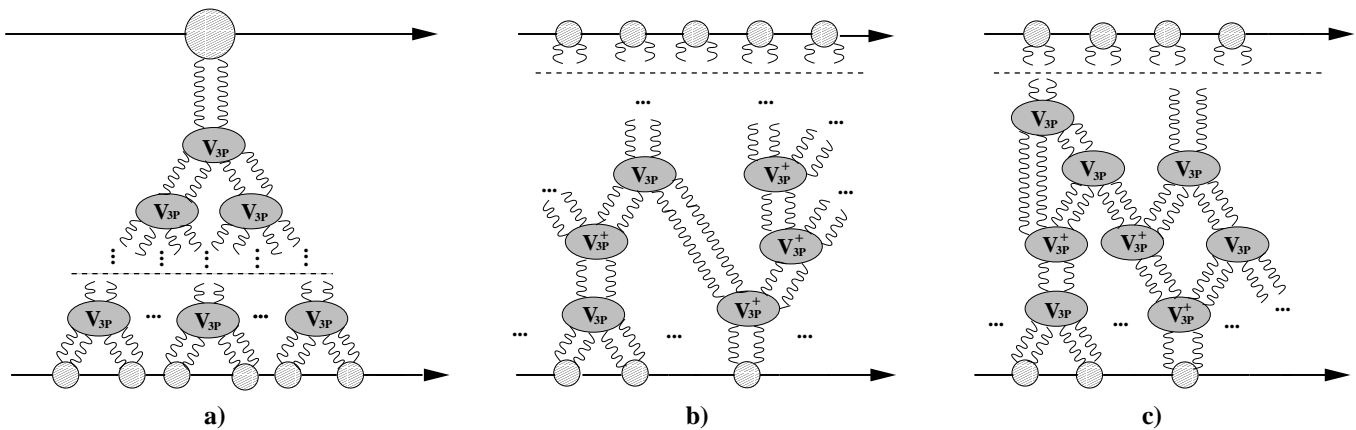,width=180mm} 
\end{center} 
\caption{\it 
Examples of diagrams of the effective field theory of QCD pomerons 
interacting with triple pomeron vertices:
a) a fan diagram; b) a tree diagram defining the classical limit; 
c) a diagram with quantum loops.}
\label{diag1}
\end{figure}

In parallel to the theoretical efforts to determine the deep fundamental 
structure of the effective field theory for high energy scattering, more 
phenomenological studies of the unitarity effects have been carried out. 
Probably, a pair of the most fruitful (and entangled) concepts of the 
last decade were the Balitsky-Kovchegov (BK) evolution 
equation~\cite{balitsky,kov1,kov2}, and 
the saturation model proposed by Golec-Biernat and W\"{u}sthoff 
(GBW)~\cite{gbw1,gbw2}. The BK equation was derived in the CGC formulation 
and in the framework of QCD-RFT it can be viewed as a resummation of 
the BFKL pomeron fan diagrams of the type depicted in \fig{diag1}a.
An important feature of the BK equation is 
its simplicity -- absence of the vertex for pomeron splitting removes 
all quantum loops of the complete theory. Therefore, a classical treatment
of PFT is exact in the case of the BK equation.

Numerous explicit solutions~\cite{bksols,bdepkov,jimsol} and 
semi-analytical analyzes of the BK equation were 
presented~\cite{bksols,bksemi1,bksemi2} and applications were developed 
for the deep inelastic scattering (DIS) of nuclei and off the 
nucleon~\cite{bk-pheno,kks,iim}.
Probably, the most remarkable features of the BK equation is generation
of the {\em saturation scale} $Q_s(y)$, increasing exponentially  with the 
available rapidity $y$: $Q_s(y) \sim \exp(\lambda y)$, and the 
{\em geometric scaling}~\cite{scaling,barlev,traveling}. 
Both the features of the BK equation are strongly favoured by the 
experimental data for $\sigma(\gamma^*(Q^2) p)$ down to $Q^2=0$,
the diffractive~DIS data and heavy flavour electro-production~\cite{gbw1,gbw2,gbwev}; two-photon cross sections~\cite{tkm}; exclusive photo- and 
electroproduction of vector mesons and the deeply virtual Compton 
scattering~\cite{kt,kmw}, etc.

Unfortunately, the BK equation is not sufficient to address the problem
of scattering of two similar objects, that is the symmetric situation, like
proton-proton and nucleus-nucleus collisions at RHIC and the LHC.
At present, efforts are made to understand the effective PFT beyond 
the BK limit~\cite{stochastic,braun1,braun2,braun3,absorptive,balshock}. 
The first key step is to determine properties of the theory after the 
vertex describing splitting of the pomeron to two pomerons is added to 
the BK framework. Importantly, the form of this vertex is imposed by the
form of the vertex for two pomeron merging. The reason for that is that
the two vertices differ only by the choice of direction of evolution in 
rapidity, which is arbitrary and should have no impact on physics.
This simple fact was known in the RFT since a long 
time~\cite{0dimc,0dimq,0dimi} and recently it was re-discovered in the 
CGC framework, as the self-duality of the CGC Hamiltonian~\cite{selfdual}. 
With the pomeron splitting  vertex quantum pomeron loops become possible, 
as exemplified in \fig{diag1}c, and solving the theory becomes much more 
difficult. There exists, however, an interesting limit in 
which some explicit results can be found. Namely, collision of two large 
nuclei composed of $A \gg 1$ nucleons each, can be analyzed. 
In the limit of very large $A$ the quantum pomeron loops will provide 
only subleading contribution, suppressed by powers $1/A$. 
Then, the tree topologies of the pomeron diagrams give the 
dominant contribution to the $S$-matrix, corresponding to the classical 
limit of the effective Pomeron Field Theory. A pomeron diagram that has 
the tree topology is shown in \fig{diag1}b. This concept was put forward and 
developed in a series of pioneering papers by 
Braun~\cite{braun1,braun2,braun3}, and an analogous
path was taken by Balitsky in the framework of the Color Glass 
Condensate~\cite{balshock}. In the present study we provide a complementary 
analysis to that of Braun, solving the classical equations of motion of the 
Pomeron Field Theory, and reporting observations of some new, unexpected 
features of the PFT. Among them, perhaps, the most surprising is a break-down
of the symmetry between the target and the projectile at the classical level.

The paper is organized as follows. In the next section we describe the 
formalism of the effective QCD Pomeron Field Theory. In Sec.~3 we 
provide insight into the theory coming from a toy model of the 
Reggeon Field Theory in zero transverse dimensions. In Sec.~4 solutions 
of PFT are presented. We discuss the results in Sec.~5 and conclude in Sec.~6.

%%%%%%%%%%%%%%%%%%%%%%%%%%%%%%%%%%%%%%%%

\section{Formalism}

\subsection{The Balitsky-Kovchegov equation and the triple pomeron vertex}

Scattering of the small perturbative probe off a large nucleus was
studied by Balitsky and Kovchegov~\cite{balitsky,kov1,kov2}. 
Balitsky derived the rapidity 
evolution equations describing the scattering amplitude in QCD, which 
involved an infinite tower of the correlators. This hierarchy is cut down to
the lowest correlator in the large $N_c$ limit, leading to a much simpler
and better tractable evolution equation. Such an equation was derived by
Kovchegov in the framework of the dipole model. Thus, the Balitsky-Kovchegov 
equation provides the correct limit of the scattering amplitude in QCD
for a large nucleus at the LL$s$ accuracy, and for $1/N_c \to 0$.
In the infinite momentum frame, the BK equation resums the BFKL pomeron fan
diagrams (see \fig{diag1}a), with the triple pomeron vertex equivalent to the 
Bartels vertex at the leading~$1/N_c$ accuracy. 

The BK equation was initially proposed as a non-linear evolution equation 
for the dipole scattering amplitude $N(\ytau;\rbo,\bbo)$, where 
the dipole spans the vector $\rbo$ and is located at the transverse 
position $\bbo$. Thus, the BK equation reads
 \beeq
\nonumber
\frac{\partial N(\ytau;\rbo,\bbo)}{\partial \ytau}
&=&
\overline{\alpha}_s\
(\tilde{\cal K}\otimes N)(\ytau;\rbo,\bbo)
\\
&-&
\overline{\alpha}_s\,
\int \frac{d^2\rbop}{2\pi}\,
\frac{r^2}{r^{\/\prime 2}(\rbo+\rbop)^2}\,
N\left(\ytau;
\rbo+\rbop,\bbo+\frac{\rbop}{2}\right)\;
N\left(\ytau;\rbop,\bbo+\frac{\rbo+\rbop}{2}
\right).
\label{eq:kov}
\eeeq
where $\overline{\alpha}_s=N_c \alpha_s/\pi$, and the linear term is
determined by the BFKL kernel in the position space
\begin{equation}
\\
\label{eq:bfklkernel}
(\tilde{\cal K}\otimes N)(\ytau;\rbo,\bbo)=
\int \frac{d^2\rbop}{2\pi\/ r^{\/\prime 2}}\,
\left\{
\frac{2\, r^2}{(\rbo+\rbop)^2}\,N(\ytau;\rbo+\rbop,\bbo)
-
\frac{r^2}{r^{\/\prime 2}+(\rbo+\rbop)^2}\,N(\ytau;\rbo,\bbo)
\right\}.
\\
\end{equation}
In the limit of the small scattering amplitude the quadratic term may be
neglected and the BFKL equation in the dipole picture is obtained.

The BK equation is a differentio-integral equation, for which
the integral kernel depends on two two-dimensional vectors $\rbo$ and 
$\bbo$. Numerical solution of the complete equation is possible~\cite{bdepkov} 
but cumbersome. Besides that, the treatment of the large dipoles and large
impact parameters in the BK equation is, strictly speaking, incorrect
as far as QCD is concerned. Namely, the confinement of colour is not 
accounted for, which can be seen, for instance from the conformal 
invariance of the equation. Thus, the Froissart limit for the scattering 
matrix is broken due to a power like diffusion to the large impact 
parameters~\cite{kovn1}.
Thus, in this work we follow the approximation made in  
the earlier analyzes that the dominant contribution to the scattering 
amplitude comes from perturbative dipoles for which $r \ll R$, 
where $R$ is the nucleus size. In this case the evolution may be assumed 
to be local in the transverse plane and Eq.~(\ref{eq:kov}) becomes 
independent of $\bbo$. Thus, the $\bbo$-dependence can be suppressed 
in $N(\ytau;\rbo,\bbo)$ and it enters only through the initial 
condition for the evolution equation.

For an azimuthally symmetric solution, $N(\ytau,\rbo)=N(\ytau,r)$,  
it is convenient to Fourier transform Eq.~(\ref{eq:kov}) 
to the momentum space,
\begin{equation}
\phi(k^2,\ytau)
\,=\,
\int \frac{d^2\rbo}{2\pi} \exp(-i\kbo\cdot \rbo)\,
\frac{N(\ytau,r)}{r^2}
\,=\,
\int_0^\infty
\frac{dr}{r}\, J_0(k\/r)\,N(\ytau,r),
\end{equation}
where $J_0$ is the Bessel function.
In this case the following equation is obtained
\beeq
\label{eq:newkov}
\frac{\partial \phi(\ytau,k^2)}{\partial \ytau}
\,=\,
\overline{\alpha}_s\, ({\cal K}^\prime\otimes \phi)(\ytau,k^2)
\,-\
\overline{\alpha}_s\, \phi^{\,2}(\ytau,k^2),
\eeeq
and the action of the BFKL kernel (suitably
shifted in the space of the Mellin moments in $k^2$) is given by
\begin{equation}
({\cal K}^\prime\otimes \phi)(\ytau,k^2)
\,=\,
\int_0^{\infty} \frac{da^2}{a^2}\,
\left\{
\frac{a^{2}\,\phi(\ytau,a^{2})\, -\, k^2\, \phi(\ytau,k^2)}
{|k^2\,-\, a^{2}|}
\,+\,
\frac{k^2\, \phi(\ytau,k^2)}{\sqrt{4 a^{4}\,+\,k^4}}
\right\},
\end{equation}
where $k^2$ and $a^2$ are the virtualities of the 
exchanged gluons in the BFKL ladder.

Equation \eq{eq:newkov} may be further transformed to the form dependent 
on the unintegrated gluon density in the transverse space~\cite{kks}.
One has
%\footnote{Strictly speaking, this relation between the scattering 
%amplitude of a dipole and the unintegrated gluon density is valid only if 
%the dipole is coupled to a pair of gluons, that is if no eikonalization 
%effects are included in this amplitude.}
\beq
f(\ytau,k^2) = {N_c \over 4 \alpha_s \pi^2} k^4 \nabla_k ^2 \phi(\ytau,k^2), 
\label{phi2f}
\eeq
and conversely,
\beq
\phi(\ytau,k^2) = {\pi^2 \alpha_s \over N_c} 
\int_{k^2} {da^2 \over a^4} f(\ytau,a^2) \,
\log\left( {a^2 \over k^2} \right). 
\label{f2phi}
\eeq
The unintegrated gluon density in the transverse space may be related  
in the small-$x$ limit to the collinear gluon distribution of the target~$A$
\beq
\int_{A}\! d^2 \vb\, f(\ytau,k^2,\vb) = 
{\partial xg(x,k^2)\over\partial \log k^2},  
\eeq 
where $y=\log(1/x)$ and we restored the dependence of $f(\ytau,k^2,\vb)$ 
on the tranverse position $\vb$, assuming that it is mild enough 
to ensure effective decoupling of the $\vb$~dependence from the BK 
evolution, which should hold for a large target.

After this transform is executed the BK equation reads~\cite{kks}
\[
\partial_\ytau f(\ytau,k^2) = 
{N_c \alpha_s \over \pi}\, k^2\, \int {da^2 \over a^2}
\left[
{f(a^2)-f(k^2) \over |a^2-k^2|} +
{f(k^2)\over [4a^4+k^4]^{{1\over 2}}}
\right]
\]
\beq
- {2\pi \alpha_s ^2} \; 
\left[
k^2 \int_{k^2} {da^2 \over a^4} \;  f(\ytau,a^2)
\int_{k^2} {db^2 \over b^4} \;  f(\ytau,b^2)
+ f(\ytau,k^2)\int_{k^2} {da^2 \over a^4} \; 
\log\left( {a^2 \over k^2} \right) f(\ytau,a^2)\right].
\eeq
It is easy to verify that the nonlinear term describing joining of two
pomerons, $(f,f) \to f$ is generated from the amplitude of the 
Bartels triple pomeron vertex (in the forward limit),
\[
(\fda|  V_{3P} | f \otimes f)  =  -{2\pi \alpha_s ^2} \; 
\int {da^2 \over a^4} \,a^2 \fda(\ytau,a^2)
\int_{a^2} {db^2 \over b^4} \;  f(\ytau,b^2)
\int_{a^2} {dc^2 \over c^4} \; f(\ytau,c^2)
\]
\beq
- {2\pi \alpha_s ^2} \; 
\int {da^2 \over a^4} \, \fda(\ytau,a^2) f(\ytau,a^2)
\int_{a^2} {db^2 \over b^4} \; \log\left( {b^2 \over a^2} \right)
f(\ytau,b^2).
\eeq
by functional differentiation with respect to the pomeron field 
$\fda(\ytau,k^2)$. The details of the complete
effective action for interacting pomerons are given in the next 
section.

\subsection{Effective action and the self-duality}

Following Braun~\cite{braun1,braun2,braun3}, 
we shall construct an effective action to represent both 
pomeron merging and splitting. We wish to study the minimal extension of 
the BK equation, thus we neglect vertices at which more than three pomerons 
meet. It is clear that the vertex for pomeron splitting must be the same 
as the vertex for pomeron merging. The reason is that in order to distinguish
merging of the pomerons from a splitting of a pomeron we have to specify the 
direction of the evolution in rapidity. This choice is, however, 
completely arbitrary and the form the action should not depend on it. 
More precisely, the inversion of the direction of evolution, $y \to -y$ 
results with the interchange of the outgoing ($\fda$) and incoming 
($f$) pomeron fields. In result, the effective action is invariant 
under the transform,
\beq
f \leftrightarrow \fda, \qquad \ytau \to -\ytau,
\label{ptsym}
\eeq 
and the form of the splitting vertex 
$(\fda\otimes \fda |  V^\dagger_{3P} | f )$ is given by 
$(\fda|  V_{3P} | f \otimes f)^\dagger$.

\begin{figure}[t]
\begin{center}
\psfig{file=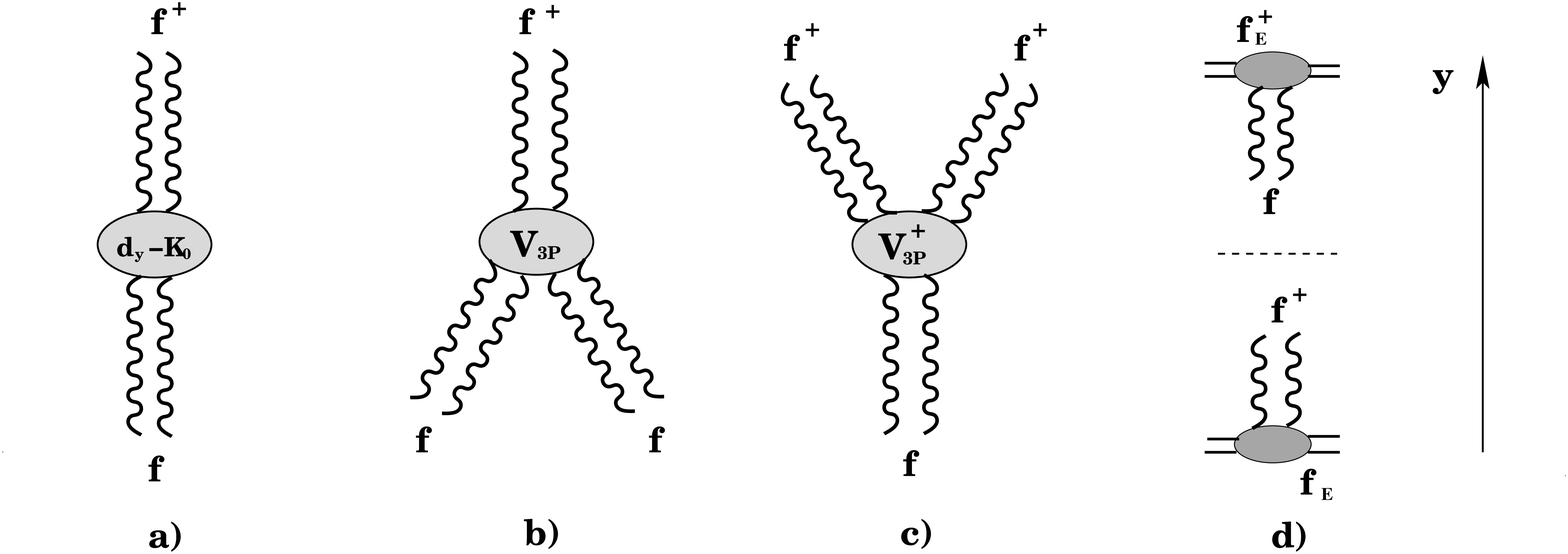,width=150mm} 
\end{center} 
\caption{\it 
Elements of the effective action: a) the BFKL pomeron propagator; b) the merging vertex;
c) the splitting vertex and d) the external sources of the fields. The arrow indicates
the direction of evolution.}
\label{diag2}
\end{figure}

Thus, in what follows we shall write the effective action for the interacting 
pomeron system based on the following principles:
\begin{enumerate}
\item Free propagation of the pomeron fields is described by the forward 
BFKL equation.
\item Pomerons interact only through triple pomeron vertices. 
\item The vertex for merging of the pomerons is the Bartels triple pomeron 
vertex at large $N_c$ and in the forward limit, equivalent to the vertex 
that generates the Balitsky-Kovchegov equation.
\item Splitting and merging of the pomerons are described by 
identical vertices.
\item The fields do not depend on the transverse position.  
\end{enumerate}
The elements of the action are graphically represented in \fig{diag2}.
The above assumptions lead to a unique form of the action,
\beq
{\cal A}[f,\fda;Y] = 
{4\pi^3 \over N_c^2 - 1} \, \int_0 ^Y d\ytau\; \left\{
{\cal L}_0[f,\fda] +  
{\cal L}_3[f,\fda] + 
{\cal L}_3 ^{\dagger}[f,\fda] +
{\cal L}_E[f,\fda] \right\},
\label{eq:aeff}
\eeq
where the Lagrange function for the free propagation reads
\[
{\cal L}_0[f,\fda] = 
{1\over 2}
\int {da^2 \over a^4} \,
\left[
f(\ytau,a^2) \partial_\ytau \fda(\ytau,a^2) -
\fda(\ytau,a^2) \partial_\ytau f(\ytau,a^2) 
\right]
\]
\beq
+ \int {da^2 \over a^4} \,
\int {db^2 \over b^4} \; 
\fda(\ytau,a^2) \kbf(a^2,b^2) f(\ytau,b^2),
\eeq
and ${\cal K}_0$ is the amputated forward BFKL kernel given by
\beq
\int {db^2 \over b^4}\kbf(a^2,b^2)f(b^2) = 
{N_c \alpha_s \over \pi}\, a^2\,
\int {db^2 \over b^2}
\left[
{f(b^2)-f(a^2) \over |b^2-a^2|} +
{f(a^2)\over [4b^4+a^4]^{{1\over 2}}}
\right].
\eeq
The Lagrange function describing merging of two pomerons takes the form 
\[
{\cal L}_3[f,\fda] =  
- {2\pi \alpha_s ^2} \; 
\int {da^2 \over a^4} \,a^2 \fda(\ytau,a^2)
\int_{a^2} {db^2 \over b^4} \;  f(\ytau,b^2)
\int_{a^2} {dc^2 \over c^4} \; f(\ytau,c^2)
\]
\beq
- {2\pi \alpha_s ^2} \; 
\int {da^2 \over a^4} \, \fda(\ytau,a^2) f(\ytau,a^2)
\int_{a^2} {db^2 \over b^4} \; \log\left( {b^2 \over a^2} \right)
f(\ytau,b^2).
\eeq
Splitting of a pomeron contributes with
\[
{\cal L}^{\dagger}_3[f,\fda] =  
-{2\pi \alpha_s ^2} \; 
\int {da^2 \over a^4} \,a^2 f(\ytau,a^2)
\int_{a^2} {db^2 \over b^4} \;  \fda(\ytau,b^2)
\int_{a^2} {dc^2 \over c^4} \; \fda(\ytau,c^2)
\]
\beq
-  {2\pi \alpha_s ^2} \; 
\int {da^2 \over a^4} \,f(\ytau,a^2) \, \fda(\ytau,a^2) \;
\int_{a^2} {db^2 \over b^4} \; \log\left( {b^2 \over a^2} \right)
\fda(\ytau,b^2).
\eeq
The coupling of the pomerons to the external sources is represented by
\beq
{\cal L}_E [f,\fda] =  
\int {da^2 \over a^4}\, \left[ 
\fda_E(\ytau,a^2) f(\ytau,a^2) + \fda(\ytau,a^2) f_E(\ytau,a^2)
\right],
\eeq
where the external sources will be assumed to be localized in
rapidity,
\beq
f_E(\ytau,a^2) = f_A(a^2) \delta(\ytau), \qquad \qquad
\fda_E(\ytau,a^2) = \fda_B(a^2) \delta(\ytau-Y).
\eeq
Clearly, $f_A$ represents the amplitude of emission of the pomeron
described by the field $f(\ytau,k^2)$ from the source at $\ytau = 0$ and 
$\fda_B$ is the coupling of $f^{\dagger}(\ytau,k^2)$ to an external source at
$\ytau = Y$.

Let us stress that the treatment of the transverse position is missing and 
thus the evaluation of the quantum loops of the complete theory cannot be 
performed with this action. Nevertheless, the action treated in the 
semi-classical framework may be used to approximately resum the BFKL 
pomeron tree diagrams (see \fig{diag1}b) in scattering of two large objects, 
for instance of two nuclei. For a scattering in which the 
projectile and the target have sizes much larger than the typical momenta in
the QCD pomeron, the momentum transfer of the pomeron line originating from
the external particles is bounded to be small by the form-factors of the 
sources and may be therefore neglected. In the diagrams without closed pomeron
loops the constraint imposed on momentum transfer of the external lines 
propagates and extends to all pomeron lines. 

In fact, a complete action that properly represents the degrees of freedom
corresponding to the momentum transfer (or equivalently to the transverse positions of the pomerons), that are missing in the action \eq{eq:aeff}, 
was proposed by Braun~\cite{braun3}. 
It is also straightforward to 
write down an analogous action in the present formulation. 
That complete action, however, leads to the same dynamics of 
a scattering of two large objects in the semi-classical limit, the problem 
that we address in this work. Therefore, we restrict ourselves to 
the simplified effective action given by \eq{eq:aeff}.

Let us return to the symmetry of the action defined by Eq.~\eq{ptsym}. 
This symmetry causes the action to be self-dual. 
Indeed, after integration by parts of the ``time derivative'' part of the 
action
\beq 
\int_0 ^Y dy \,
{1\over 2}
\left[
f(\ytau,a^2) \partial_\ytau \fda(\ytau,a^2) -
\fda(\ytau,a^2) \partial_\ytau f(\ytau,a^2) 
\right]
\to 
\int_0 ^Y dy \,
\left[
-\fda(\ytau,a^2) \partial_\ytau f(\ytau,a^2) 
\right] + (\ldots)
\eeq
where $\, (\ldots) \,$ denote the boundary terms, one gets that 
\beq
{\delta {\cal L}[f,\fda] \over \delta (\partial_\ytau f(\ytau,k^2))}
= -{1 \over k^4} \fda(\ytau,k^2).
\label{eq:canmom}
\eeq 
This means, that the field $\fda(y,k^2)$ is the canonical conjugate of $f(y,k^2)$, 
up to the factor of ${1/k^4}$ which can be easily absorbed into the field definitions
and trivial complex phase factors. After invoking the symmetry~\eq{ptsym} we conclude
that the bulk part of the action \eq{eq:aeff} may be rewritten in the self-dual form. 
The symmetry of the action \eq{eq:aeff} may be completed by assuming the symmetric 
external sources, that enter ${\cal L}_E$. Then, one expects the solution of the 
field equations $\{f,\fda\}$ to be also symmetric 
\beq
f(y,k^2) = \fda(Y-y,k^2). 
\label{eq:tarpro}
\eeq 
In what follows, we shall refer to this as to the {\em projectile-target symmetry}.

\subsection{Equations of motion}

Let us list the functional derivatives of the action of the 
effective Pomeron Field Theory \eq{eq:aeff} with respect to
$\fda(\ytau,k^2)$:
\beq
{\delta {\cal L}_0[f,\fda] \over \delta (\partial_\ytau\fda(\ytau,k^2))}
= {1\over 2}{1 \over k^4} f(\ytau,k^2);
\label{eq:diff1}
\eeq

\beq
{\delta {\cal L}_0[f,\fda] \over \delta \fda(\ytau,k^2)}
= -{1 \over k^4} 
\left[ {1\over 2} \partial_\ytau f(\ytau,k^2) - 
\int {db^2 \over b^4} \kbf(k^2,b^2)  f(\ytau,b^2) \right];
\label{eq:diff2}
\eeq

\beq
{\delta {\cal L}_3[f,\fda] \over \delta \fda(\ytau,k^2)}
=  -{2\pi \alpha_s ^2} \; 
{1\over k^4} \left[
k^2 \int_{k^2} {da^2 \over a^4} \;  f(\ytau,a^2)
\int_{k^2} {db^2 \over b^4} \;  f(\ytau,b^2)
+ f(\ytau,k^2)
\int_{k^2} {da^2 \over a^4} \; \log\left( {a^2 \over k^2} \right)
f(\ytau,a^2)
\right];
\label{eq:diff3}
\eeq

\[
{\delta {\cal L}^{\dagger}_3[f,\fda] \over \delta \fda(\ytau,k^2)} 
=  -{2\pi \alpha_s ^2} \; 
{1\over k^4} \left[
2 \int_0 ^{k^2} {da^2 \over a^4} \,a^2 f(\ytau,a^2)
\int_{a^2} {db^2 \over b^4} \; \fda(\ytau,b^2) \right.
\] 
\beq
\left.
+  f(\ytau,k^2) 
\int_{k^2} {da^2 \over a^4} \; \log\left( {a^2 \over k^2} \right)
\fda(\ytau,a^2) 
+ \int_0 ^{k^2} {da^2 \over a^4} \, f(\ytau,a^2) \, \fda(\ytau,a^2) \;
 \log\left( {k^2 \over a^2} \right)
\right];
\label{eq:diff4}
\eeq

\beq
{\delta {\cal L}_E[f,\fda] \over \delta \fda(\ytau,k^2)} = 
{1 \over k^4} f_E(\ytau,k^2). 
\label{eq:diff5}
\eeq 

Analogously one computes the functional derivatives with respect to
$f(\ytau,k^2)$. The results of that procedure may be obtained by 
an interchange of $f\leftrightarrow \fda$ in equations 
(\ref{eq:diff1}--\ref{eq:diff5}) and changing the sign of $f(\ytau, k^2)$ 
in \eq{eq:diff1} and of the $\partial_\ytau f(\ytau, k^2)$ in \eq{eq:diff2}. 
Thus, one obtains the following equations of motion,
\[
\partial_\ytau f(\ytau,k^2) = 
{N_c \alpha_s \over \pi} \, k^2\,\int {da^2 \over a^2}
\left[
{f(a^2)-f(k^2) \over |a^2-k^2|} +
{f(k^2)\over [4a^4+k^4]^{{1\over 2}}}
\right]
\]
\[
- {2\pi \alpha_s ^2} \; 
\left[
k^2 \int_{k^2} {da^2 \over a^4} \;  f(\ytau,a^2)
\int_{k^2} {db^2 \over b^4} \;  f(\ytau,b^2)
+ f(\ytau,k^2)\int_{k^2} {da^2 \over a^4} \; \log\left( {a^2 \over k^2} \right)
f(\ytau,a^2)\right]
\]
\[
-  {2\pi \alpha_s ^2} \; 
\left[
2 \int_0 ^{k^2} {da^2 \over a^4} \,a^2 f(\ytau,a^2)
\int_{a^2} {db^2 \over b^4} \; \fda(\ytau,b^2)
+  f(\ytau,k^2) 
\int_{k^2} {da^2 \over a^4} \; \log\left( {a^2 \over k^2} \right)
\fda(\ytau,a^2) \right]
\]
\beq 
- {2\pi \alpha_s ^2} \; 
\int_0 ^{k^2} {da^2 \over a^4} \, f(\ytau,a^2) \, \fda(\ytau,a^2) \;
 \log\left( {k^2 \over a^2} \right),
\label{evolf}
\eeq
%%%%%%%%%%%
and
%%%%%%%%%%
\[
-\partial_\ytau \fda(\ytau,k^2) = 
{N_c \alpha_s \over \pi}\, k^2\, \int {da^2 \over a^2}
\left[
{\fda(a^2)-\fda(k^2) \over |a^2-k^2|} +
{\fda(k^2)\over [4a^4+k^4]^{{1\over 2}}}
\right]
\]
\[
- {2\pi \alpha_s ^2} \; 
\left[
k^2 \int_{k^2} {da^2 \over a^4} \;  \fda(\ytau,a^2)
\int_{k^2} {db^2 \over b^4} \;  \fda(\ytau,b^2)
+ \fda(\ytau,k^2)\int_{k^2} {da^2 \over a^4} 
\; \log\left( {a^2 \over k^2} \right) \fda(\ytau,a^2)\right]
\]
\[
- {2\pi \alpha_s ^2} \; 
\left[
2 \int_0 ^{k^2} {da^2 \over a^4} \,a^2 \fda(\ytau,a^2)
\int_{a^2} {db^2 \over b^4} \; f(\ytau,b^2)
+  \fda(\ytau,k^2) 
\int_{k^2} {da^2 \over a^4} \; \log\left( {a^2 \over k^2} \right)
f(\ytau,a^2) \right]
\]
\beq 
-  {2\pi \alpha_s ^2} \; 
\int_0 ^{k^2} {da^2 \over a^4} \, \fda(\ytau,a^2) \, f(\ytau,a^2) \;
 \log\left( {k^2 \over a^2} \right),
\label{evolfd}
\eeq
with the two-point boundary conditions,
\beq
f(\ytau=0,k^2) = f_A(k^2), \qquad\qquad  \fda(\ytau=Y,k^2) = \fda_B(k^2).
\label{boundary}
\eeq

The equations are equivalent to the equations derived by Braun~\cite{braun1}, 
although they are formulated using other variables. 
Therefore we shall refer to equations (\ref{evolf}), (\ref{evolfd})
as to the {\em Braun equations}.
Apparently, the present formulation is more complicated and less convenient 
than the original one. Still, it might be advantageous to use the present 
form. The reason is that the interpretation of the  degrees of 
freedom that we use is straightforward in terms of perturbative QCD
in the momentum space; the basic physical objects: the unintegrated gluon 
and the triple pomeron vertex in the momentum space are represented 
in a transparent way. Using the present form it should be also relatively 
simple to account for non-leading corrections to the BFKL kernel, 
as it was done for the BK equation~\cite{kks}.

\subsection{The $S$-matrix}

Solutions to classical equations of motions for the pomeron fields 
may be used to determine the $S$-matrix for the high energy scattering 
in the semi-classical approximation. In order to do that, however, 
the dependence of the problem on the transverse position has to be taken 
into account.

First, let us consider the general case, in which the action has 
the complete dependence on the transverse position. Thus, the pomeron 
fields depend on the position $\vb_1$ and $\vb_2$ with respect to the 
center of the projectile and the target, correspondingly: 
$f(y,k^2) \to \tilde f(y,k^2, \vb_1)$ 
and $\fda(y,k^2) \to \tilde\fda(y,k^2,\vb_2)$. 
Suppose that we know the solutions to the Braun equations with the full 
dependence of the transverse position 
for a given impact parameter $\vb$ of the collision. 
Then, the complete action may be evaluated for such a solution 
by performing integrations over transverse positions of all the fields, 
with the weights provided by the $\vb$-dependent Lagrangian density.

In this paper we do not attempt to resolve the complex dynamics of the 
fields in the transverse plane. Therefore we should apply an approximate 
treatment, in the spirit of the original work of Kovchegov and following 
the initial Braun proposal. 
Those authors assumed that the sources of pomeron fields were 
large nuclei. Those objects are much larger than typical
pomeron sizes, defined as inverse of the typical gluon virtualities in the 
pomeron. Therefore, for the bulk of interactions, an approximate 
translational invariance in the transverse space holds. This is not true 
only at the nucleus boundary, which gives, however, only a subleading 
contribution to the scattering amplitude. The simplest way to approximately
account for this situation is to assume that the action is local in the 
transverse position. Then, it is enough to solve the Braun equations with 
input conditions dependent on the transverse position,  
$\tilde f_A(k^2,\vb_1)$ and $\tilde\fda_B(k^2,\vb_1-\vb)$ 
at given impact parameter vector $\vb$. In writing so, we assume that
the initial condition $\tilde f_A(k^2,\vb_1)$ is centered at $b_1=0$ and 
the distribution $\tilde\fda_B(k^2,\vb_1-\vb)$ develops around the point
whose position is given by $\vb$. For instance, for a collision of two
cylindrical nuclei with the same radius $R$, one has: 
$\tilde f_A(k^2,\vb_1) =   f_A(k^2) \Theta(R-b_1)$ and
$\tilde \fda_B(k^2,\vb_2) =   \fda_B(k^2) \Theta(R-b_2)$.

Thus, assuming locality of the evolution in the transverse space
the complete action takes the form
\beq
\tilde{\cal A}[\tilde{f},\tilde{f}^{\dag};Y,\vb]\,=\,
\int\,d^2 \vb_1\;{\cal A}[\tilde{f}(y,k^2,\vb_1),
\tilde{f}^{\dag}(y,k^2,\vb-\vb_1);Y],
\eeq
where the equations of motion may be employed to obtain
\beq 
{\cal A}[\tilde{f},\tilde\fda;Y] = 
{1\over2} \int_{0-} ^{Y^+} dy 
\left\{ {\cal L}_E[\tilde f,\tilde\fda] 
-{\cal L}_3[\tilde f,\tilde\fda]   
-{\cal L}^\dagger _3[\tilde f,\tilde\fda] \right\},  
\eeq
leading, in the semi-classical approximation, to the $S$-matrix
\beq
S(Y,\vb)\,=\,\exp\{-\tilde{\cal A}[\tilde{f},\tilde{f}^{\dag};Y,\vb]\}\,.
\eeq
Some more details and subtleties of this approximation will be discussed 
in the next section.

%%%%%%%%%%%%%%%%%%%%%%%%%%%%%%%%%%%%%%%%%

\section{Toy model -- Reggeon Field Theory in zero transverse dimensions}
\label{sec:rft0}

\subsection{Formulation}

As a constructive example of a possible qualitative behavior of the
interacting pomeron system we consider a zero dimensional toy model 
of interacting pomerons, the, so called, Reggeon Field Theory in zero 
transverse dimensions (RFT-0). 
The model of RFT-0 was formulated and studied in depth long time ago, 
see e.g.\ \cite{0dimc,0dimq} and recently it has enjoyed a revived 
interest~\cite{0dimnew,0dimkl}. 
In fact, it turns out that RFT-0 in the weak coupling regime exhibits
some generic features which seem to be present also for interacting QCD 
pomerons. Therefore, this much simpler model may be used to provide 
some insight into the complex dynamics of QCD Pomeron Field Theory.

This model is determined by the action
\beq
{\cal A}_{RFT-0}[q(y),p(y);Y] = \int_0 ^Y dy \, {\cal L}_{\mathrm{RFT-0}}.
\label{arft}
\eeq 
with the Lagrangian: 
\beq\label{RFT1}
{\cal L}_{\mathrm{RFT-0}}\,=\,\frac{1}{2}\,q\,\partial_y {p}\,-
\frac{1}{2}\,p\,\partial_y {q}\,+
\,\mu\,q\,p\,-\,\lambda\,q\,(q\,+\,p)\,p\,
+\,p(y)\,q_0(y)\,+\,p_0(y)\,q(y)\,\,,
\eeq
where $\mu$ is the intercept of the pomeron, $\lambda$ is the triple pomeron 
coupling and $\{q,p\}$ are (up to complex phase factors) Gribov 
fields depending only on rapidity and responsible for the creation 
and annihilations of pomerons.
They correspond to $f$ and $\fda$ of the BFKL Pomeron Field Theory. 
The functions $\,q_0(y)\,$ and $\,p_0(y)\,$  are the external sources 
of the $q$ and $p$ fields respectively. 
In analogy to the assumptions of the previous section
we shall consider a scattering process at rapidity $Y$ with the 
source terms assumed to take the form:
\beq
q_0(y)\, = \, g_1 \delta(y), \qquad p_0(y)\, = \, g_2 \delta(y-Y). 
\eeq
Note, that the action is invariant under the duality transformation 
\beq
p \leftrightarrow q\qquad \mbox{and}\qquad y \to Y-y
\label{self0}
\eeq
for symmetric boundary conditions $g_1=g_2$ (obviously, the bulk action 
is invariant for any external couplings).

Dynamics of the system defined by the Lagrangian \eq{RFT1} was intensively
investigated both in the complete quantum framework~\cite{0dimq} and in 
the semi-classical approximation~\cite{0dimc}. 
Here, we focus on the latter treatment in order to match the approximations 
which we use in the description of the BFKL Pomeron Field Theory.  
Thus, the quantum evolution of the system may be represented by the 
$S$-matrix expressed using the path integral
\beq
S(Y;g_1,g_2) = \int [Dq\,Dp] 
\exp \left\{-{\cal A}_{RFT-0}[q(y),p(y);Y] \right\},  
\label{s-path}
\eeq 
where the probe trajectories obey $q(0) = 0$ and $p(Y)=0$, as the initial 
conditions are absorbed in the action. In the semi-classical limit, 
the dominant contribution to the path integral comes form the 
classical trajectories $\{\bar q_\alpha, \bar p_\alpha\}$ 
for which the action is stationary,
\beq
S(Y;g_1,g_2) \simeq \sum_{\alpha} \Delta_\alpha
\exp \left\{-{\cal A}_{RFT-0}[\bar q_\alpha, \bar p_\alpha;Y] \right\},  
\label{s-semi}
\eeq
where $\Delta_\alpha$ represent the quantum weights of subsequent classical 
trajectories, which at the leading approximation come from resummation of 
Gaussian quantum fluctuations around the classical trajectory. 
Note, that the system evolves in rapidity which is formally equivalent 
to an evolution in the Euclidean time, thus the $S$-matrix is dominated by 
classical trajectories with the minimal value of the action. In is important
to stress that the action of RFT-0 exhibits the feature of self-duality 
(or projectile-target symmetry) as its PFT counterpart.

\subsection{Solutions: spontaneous breaking of projectile-target symmetry}
\label{sec:ssb0}

The extremal value of the action is reached for classical trajectories
$\{q,p\}$ which obey the equations of motion,
\begin{eqnarray}\label{RFT2}
\,&\,&\,\partial_y q\,=\,\mu\,q\,-\,\lambda\,q^2\,-\,2\,\lambda\,q\,p\,\\
\,&\,&\,-\partial_y p\,=\,\mu\,p\,-\,\lambda\,p^2\,-\,2\,\lambda\,q\,p\,
\end{eqnarray}
with the two-side boundary condition
\beq
q(0)\,=\,g_1\,, \qquad\,\,p(Y)\,=\,g_2.
\eeq
For $g_1 < \mu / \lambda$ and $g_2 < \mu / \lambda$ the classical 
trajectories are confined to a triangle in the phase space spanned by
points with $(p,q)$ coordinates: $(0,0)$, 
$(0,\mu/\lambda)$ and $(\mu/\lambda,0)$.
These boundary conditions permit for existence of multiple solutions provided 
that rapidity $Y$ is large enough. Thus, for $Y$ smaller than a critical 
value~$Y_c$ (depending on $g_1$, $g_2$, $\lambda$ and $\mu$) there exists 
a unique solution to the classical problem, 
$\{\bar q_1(y;g_1,g_2), \bar p_1(y; g_1, g_2)\}$. 
In the case of $g_1 = g_2=g$ the solution preserves the symmetry between the 
target and the projectile,  
\beq
\label{sym0}
\bar q_1(y;g,g) = \bar p_1(Y-y;g,g).
\eeq

\begin{figure}[h]
\begin{center}
\begin{tabular}{cc}
\psfig{file=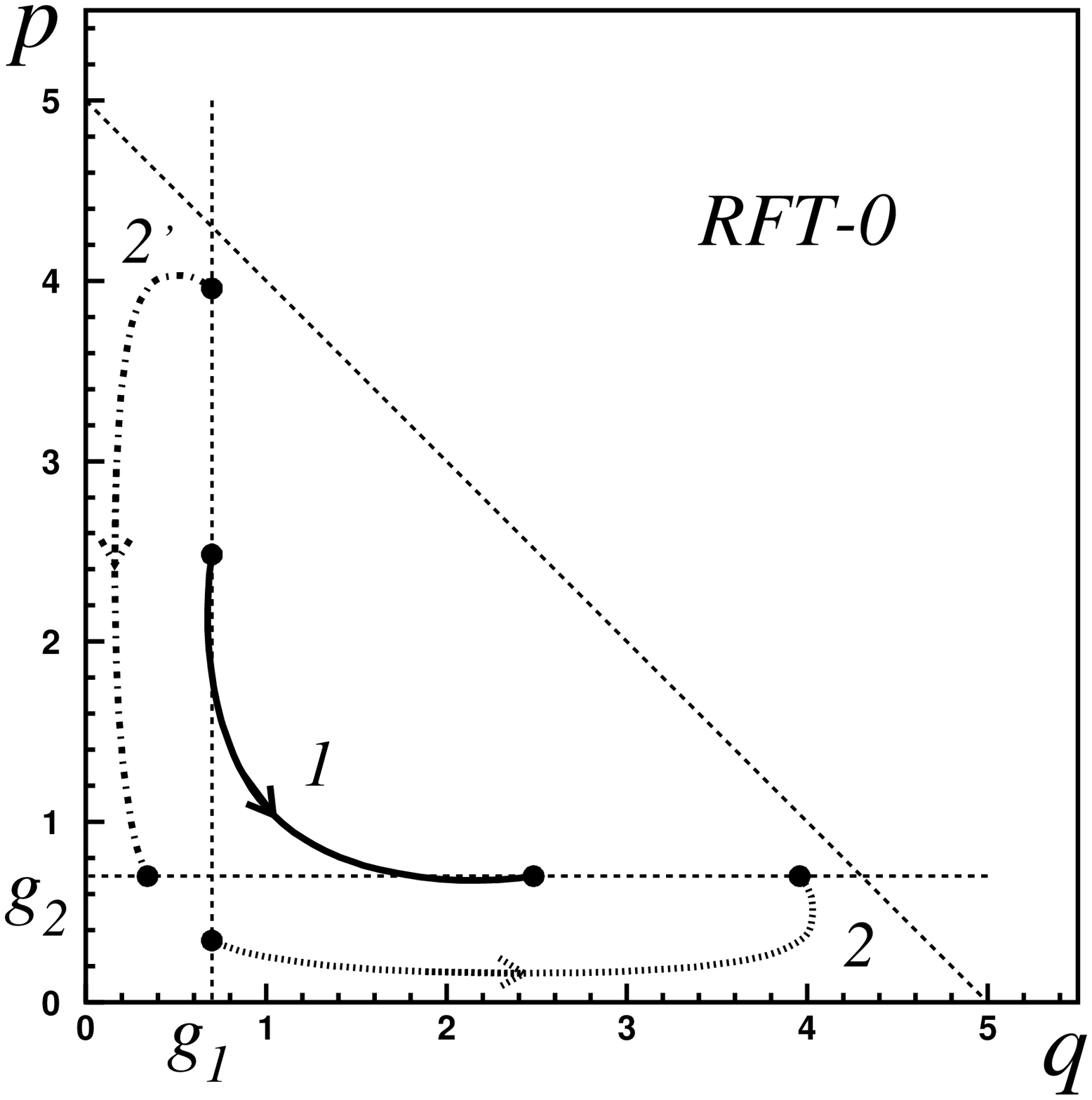,width=80mm} 
&
\psfig{file=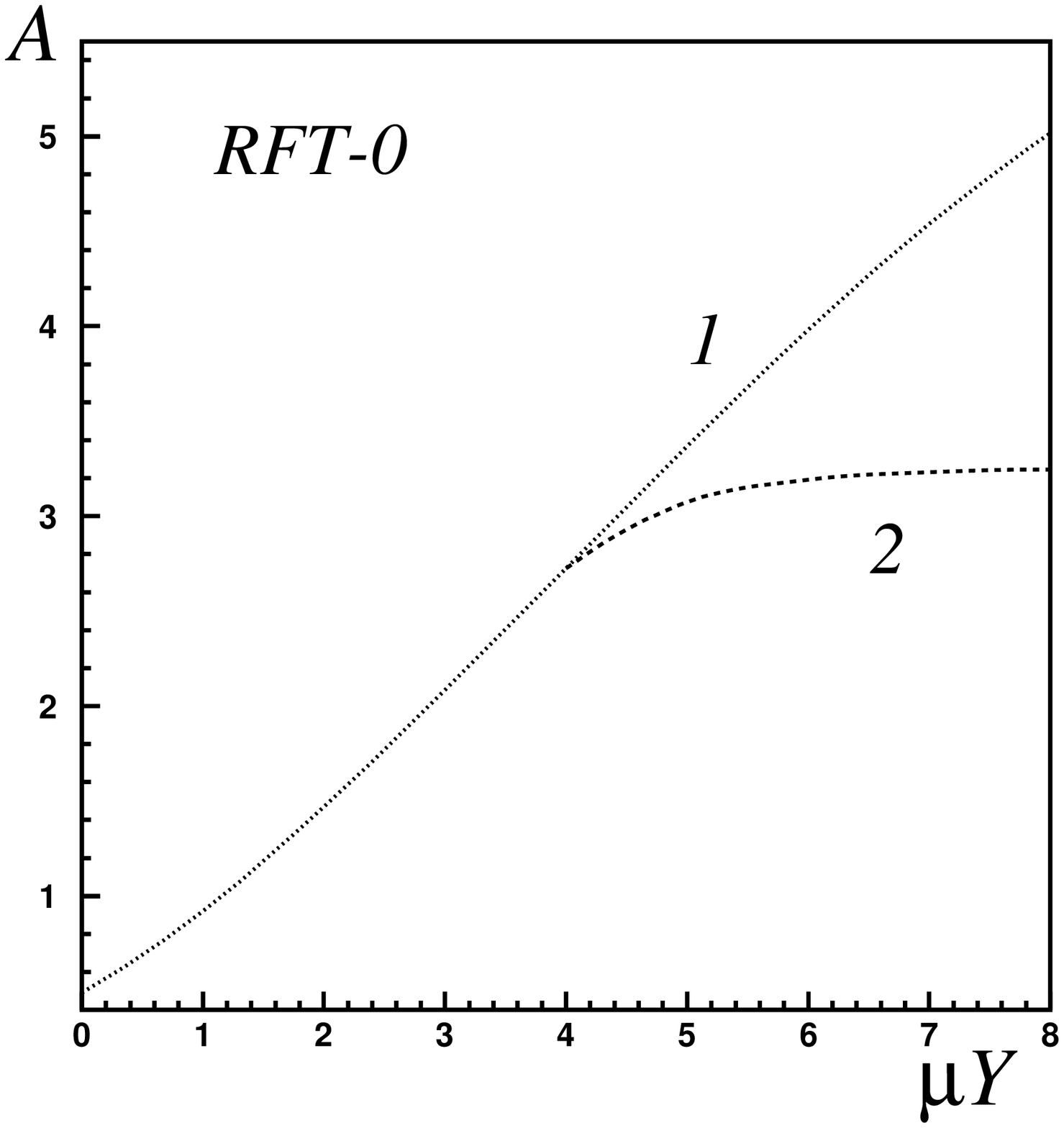,width=80mm} \\
\fig{RFT7}-a  & \fig{RFT7}-b
\end{tabular}
\end{center} 
\caption{\it Classical solutions of the RFT-0: 
a) the $\{ q,p\}$ trajectories for $Y>Y_c$; 
b) value of the action ${\cal A}_{RFT-0}[\bar q(y;g,g),\bar p(y;g,g);Y]$
for the symmetric solution (dotted line) and the asymmetric solution
(dashed line) as a function of scaled rapidity $\mu Y$.}
\label{RFT7}
\end{figure}

This simple picture changes at $Y=Y_c$. In this point two more solutions 
$\{\bar q_2(y;g,g), \bar p_2(y;g,g)\}$ and
$\{\bar q_2 '(y;g,g), \bar p_2 '(y;g,g)\}$
become possible which do not inherit the symmetry between the target and the
projectile embedded in the action and the boundary conditions,
\beq
\label{nsym0}
\bar q_2(y;g,g) \neq \bar p_2(Y-y;g,g), \quad \mbox{and} 
\quad  q'_2(y,g,g) \neq \bar p'_2(Y-y;g,g).
\eeq
An example of the solutions is given in \fig{RFT7}a plotted in the phase 
space $\{p,q\}$. The parameters of the model were chosen to be 
$\mu / \lambda  = 5$, $g_1 = g_2 = 0.7$, and $\mu Y = 8$.  
For this rapidity, one finds the symmetric trajectory~1 and two 
asymmetric trajectories:~2 and~2'. 
At yet larger values of rapidity~$Y$ more solutions are possible, 
corresponding to cycles in the phase space and giving larger values 
of the action, so we neglect those cycles in the present analysis.

The value of the action corresponding to trajectories~1 and~2 
is plotted in \fig{RFT7}b as a function of the total rescaled rapidity 
$\mu Y$. Note, that trajectory~2 is only possible for $Y > Y_c$, and the 
critical rapidity $Y_c \simeq 4$ for our choice of parameters. Clearly, the
value of the action is smaller for the asymmetric trajectories, therefore 
the asymmetric trajectories are expected to dominate the Euclidean path
integral defining the scattering amplitude at large rapidities. 
At $Y=Y_c$, however, the action of the asymmetric trajectory joins 
smoothly the action of the symmetric trajectory. 
Thus, one concludes that at the transition region of $Y \simeq Y_c$
the contribution of trajectory~1 to the scattering amplitude should be also 
included.

Note, that the emergence of the dominant asymmetric solutions may be
interpreted in terms of {\em spontaneous breaking of a discrete symmetry}
of the action. The symmetry between the projectile and the target 
(leading to the self-duality of the action) is built in the action 
(\ref{arft}) and in the boundary conditions. 
Clearly, this is a discrete symmetry. The dominant solutions of the 
equations of motion, however, are not symmetric. Thus, the symmetry is 
spontaneously broken. This is possible, as the boundary conditions are 
defined at two points of rapidity and the classical solutions need not 
be unique. As usual, however, the symmetry still holds for the full 
set of solutions,
\beq
\bar q'_2(y;g,g) = \bar p_2(Y-y;g,g)\quad \mbox{and} \quad 
\bar p'_2(y;g,g) = \bar q_2(Y-y;g,g).
\eeq
This means that the under the duality transformation (\ref{self0}) 
each solution is transformed into itself (solution~1) or into 
another solution (solutions~2 and~$2'$), and
the full set of solutions $\{\bar p,\bar q\}$ is 
invariant under the duality transformation of $p \leftrightarrow q$
and $y \to Y-y$.

\begin{figure}[h]
\begin{center}
\psfig{file=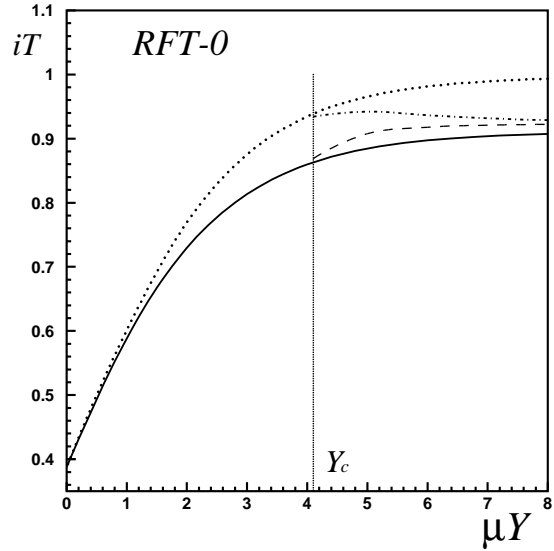,width=80mm} 
\end{center} 
\caption{\it Scattering amplitude $iT = 1-S$ from RFT-0 for 
$\,g_1\,=\,g_2\,=\,0.7\,$ at $\mu / \lambda\,=\,5\,$ as a function of 
rescaled rapidity $\mu Y$ in various approximations:
dotted line represents amplitude obtained from the symmetric solution~1, 
dashed line correspond to the pair of asymmetric solutions $\{2,2'\}$, 
the dash-dotted line accounts for sum of contributions from $\{1,2,2'\}$
and the solid line represents the full quantum solution of the problem.
(the plot is taken from~\cite{0dimour}).
}   
\label{RFT8}
\end{figure}

The observed phenomenon of spontaneous symmetry breaking  
occurs at the classical level.  At the quantum level, however, 
the projectile-target symmetry should hold\footnote{Due to quantum coherence
in finite quantum systems the spontaneous symmetry breaking does not happen. 
It is only possible in the thermodynamical limit when the coherence between 
the asymmetric configurations is broken.}. 
One sees it, for instance, from the form of the $S$-matrix in the 
semi-classical approximation using the three solutions $\{1,2,2'\}$. 
The calculation of the quantum weights for $Y>Y_c$ was performed 
in \cite{0dimi} leading to,
\beq
S(Y;g_1,g_2) \simeq 
-\exp \left\{-{\cal A}_{RFT-0}[\bar q_1, \bar p_1;Y]\right\}
+\exp \left\{-{\cal A}_{RFT-0}[\bar q_2, \bar p_2;Y]\right\}
+\exp \left\{-{\cal A}_{RFT-0}[\bar q'_2, \bar p'_2;Y]\right\},
\eeq
where the minus sign of the first term comes form the complex phase factors  
picked up by the trajectory~1 at the turning points. One sees that the symmetry
between the projectile and the projectile is restored for the $S$-matrix 
already at the semi-classical level, by summation over the complete set
of (asymmetric and symmetric) classical trajectories. 
It is interesting to ask, however, whether there could be some signs 
of the symmetry breaking found at the classical level, 
that would be seen for more exclusive observables, like for the 
rapidity distribution of produced particles. 
In principle, such a {\em classical measurement} should destroy the quantum 
coherence and select just one of the classical solutions. 
If this were true, it should lead to an asymmetric particle production 
between the identical projectile and target in individual events.

It is instructive to compare various approximations to the scattering
amplitude $iT = 1-S$. Thus, we plot in \fig{RFT8}a the scattering amplitude 
evaluated in the semi-classical approximation assuming that it is dominated  
by the symmetric solution~1 (which should be valid for $Y<Y_c$), 
by the pair of asymmetric solutions~$\{2,2'\}$, (which should hold for 
$Y\gg Y_c$), and including contributions to the $S$-matrix of all  
three solutions for $Y>Y_c$. 
We expect the last and the most complete choice be the most 
accurate. The comparison to the exact quantum evaluation of the 
scattering amplitude\footnote{The curve was obtained in the course of 
our ongoing study of the RFT-0~\cite{0dimour}, 
where we solve the RFT-0 
both at the classical and the quantum level. We leave the description
of the details to that paper.} (the continuous curve) to the various 
semi-classical evaluations reveals, however, that the contribution
of $\{2,2'\}$ approximates the exact answer best. It may be a coincidence
or a hint on the subtle point of how to treat contributions to the $S$-matrix 
from subleading trajectories. We leave the issue for further studies.

\subsection{Fan dominance}

%{\bf Continue here}

\begin{figure}[t]
\begin{tabular}{cc}
\psfig{file=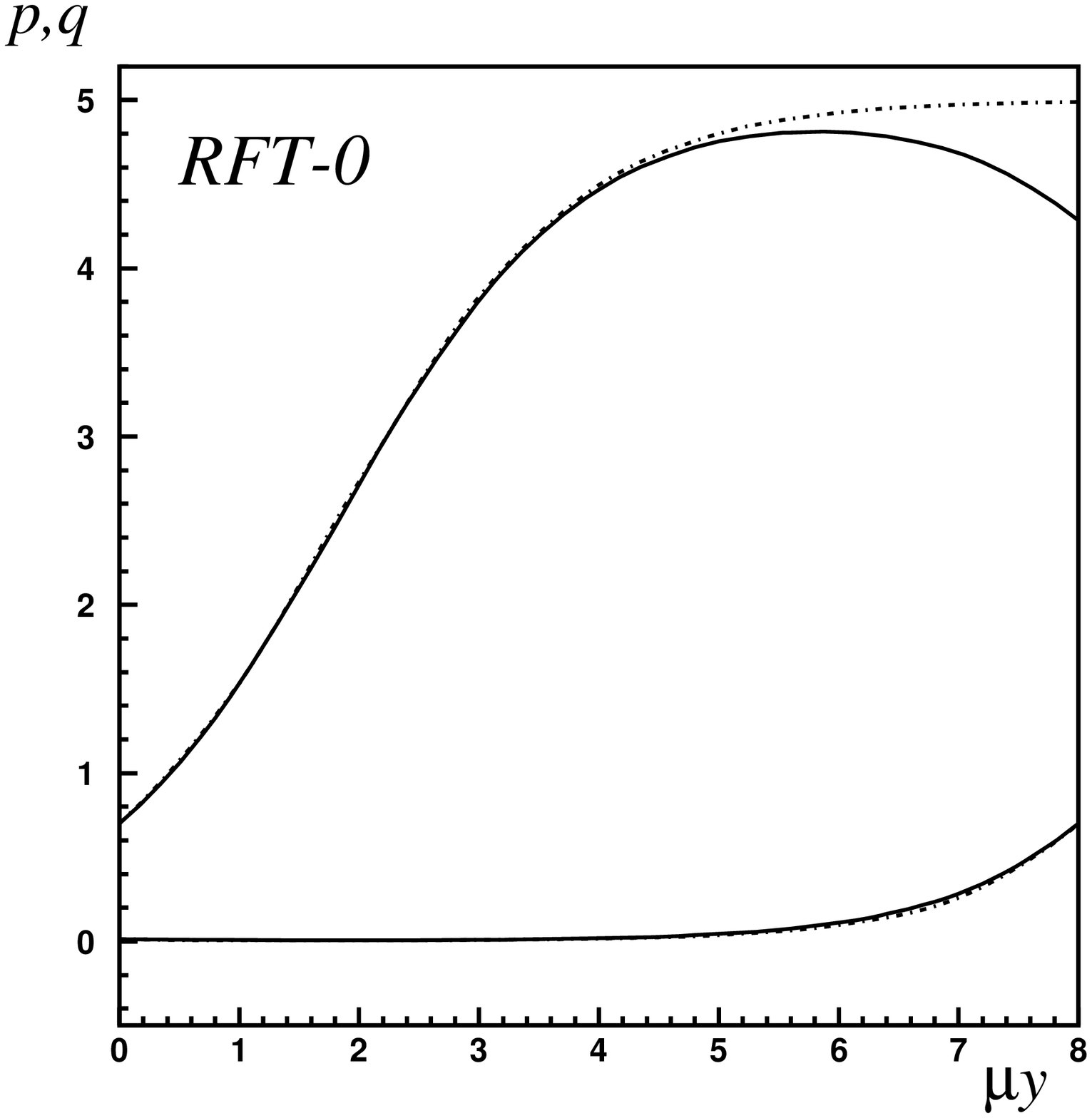,width=80mm} &
\psfig{file=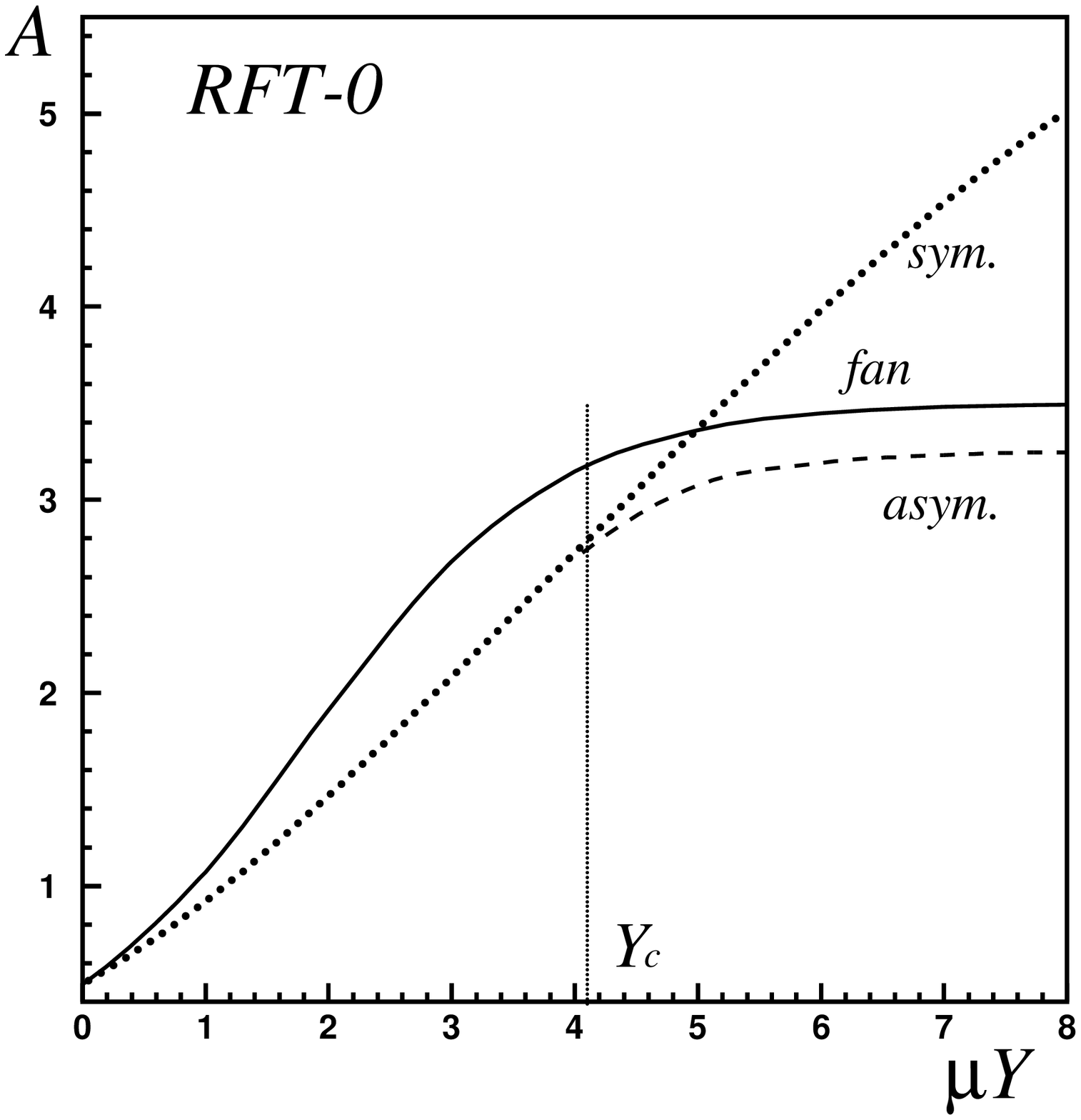,width=80mm} \\
\fig{RFT6}-a & \fig{RFT6}-b \\
\end{tabular}
\caption{\it
Comparison of the classical asymmetric solution of the full theory 
to the solution of the equation resumming the fan diagrams:
a) the trajectory: $\bar q_2(y)$ (upper solid line) and 
$\bar p_2(y)$ (lower solid line) and the trajectories of the 
fan equation $\bar q_f(y)$ and $\bar p_f(y)$ (upper and lower
dotted lines respectively) for the total rapidity $\mu Y=8$ as a 
function of $\mu y$;   
b) the action for the classical solutions: 
the asymmetric one (the dashed line) 
the symmetric one (the dotted line) versus the 
action resumming pomeron fan diagrams as a function of $\mu Y$. }
\label{RFT6}
\end{figure}

Anticipating the results of the next section, let us briefly mention an
interesting feature of asymmetric solutions for $Y$ being significantly larger 
than the critical rapidity $Y_c$ and for sources $g_1,g_2 \ll \mu / \lambda$. 
Then, one of the fields, say $q$, grows with the increasing rapidity. The 
other field, $p$, assumed to take the value of $g_2\ll \mu / \lambda$ 
at rapidity~$Y$ decreases further for decreasing $y$. 
Therefore the term involving $p^2$ in Lagrangian (\ref{RFT1}) is of 
little importance and it may be neglected. Thus, one ends up with the 
evolution equations of the system with the absent vertex for the 
pomeron splitting. This evolution equation resums the {\em fan diagrams} 
of the merging fields $q$, and it will be referred to as the {\em fan 
equation}.

We exemplify in \fig{RFT6}a the similarity of the asymmetric solution of the 
full theory $\{\bar q_2(y),\bar p_2(y)\}$ to the solution of the 
equation resumming the fan diagrams  $\{\bar q_f(y),\bar p_f(y)\}$.
The figure was obtained with $\mu/\lambda = 5$, $g_1=g_2=0.7$, and we
set $\mu Y= 8 \simeq 2\mu Y_c$.  The smaller fields $p_2(y)$ and $p_f(y)$
turned out to overlap with high accuracy. We find some noticeable difference
between  $q_2(y)$ and $q_f(y)$ only at higher values of $\mu y$. 
We checked that the asymmetric solution is closer to the ``fan equation'' 
solution if $Y$ is larger and $g_1$ and $g_2$ are smaller.
The results of evaluation of the action corresponding to  
$\{\bar q_2(y),\bar p_2(y)\}$, $\{\bar q_f(y),\bar p_f(y)\}$ and to the 
symmetric classical solution $\{\bar q_1(y),\bar p_1(y)\}$ are shown 
in \fig{RFT6}b. Again, the solution to the fan equation yields the action 
that reasonably well approximates the action 
${\cal A}_{RFT-0}(\bar q_2(y;g,g),\bar p_2(y;g,g);Y)$ 
of the dominant trajectory.  

We find this convergence of the classical system to the fan-dominated regime
to be a rather surprising effect especially in view of the fact that we 
started from completely symmetric boundary conditions. A similar observation,
however, was made in sligtly different realisation of RFT-0, with a non-zero 
four pomeron coupling~\cite{0dimkl}.
It will be very interesting to check whether 
a similar phenomenon occurs for the theory of interacting QCD pomerons.

\section{Solutions of the Braun equations}

\subsection{Parameters and the solving procedure}

Classical equations of motion \eq{evolf} and \eq{evolfd} 
for the effective pomeron fields $f(y,k^2)$ and $\fda(y,k^2)$ 
were solved numerically for various values of the total rapidity $Y$ of 
the scattering. We chose a fixed coupling constant $\alpha_s = 0.2$.
The boundary conditions were assumed to be symmetric
\beq
\label{input}
f_A(k^2) = \fda_B(k^2) = {N\over \pi R^2} {k^4 \over Q_0 ^4 + k^4}.
\eeq
The form of the input condition was inspired by the properties of 
a saturated gluon distribution in the nucleon. 
Thus we set $R^2 = 8$~GeV$^{-2}$ in order to match the preferred 
value of the nucleon size. The scale $Q_0^2 = 0.5$~GeV$^2$  
corresponds to the saturation scale for $10^{-3} < x < 10^{-2}$.
The overall normalisation factor $N=2$ so that the collinear gluon 
distribution obtained from the input is similar to the actual 
collinear gluon distribution $xg(x,Q^2)$ in the proton for 
$10^{-3} < x < 10^{-2}$ and for moderate $Q^2$.  Let us stress
that we made that choice only to pin down the physically relevant 
ranges of parameters.

The numerical solution was based on a Chebyshev interpolation method 
in the variable $\log(k^2)$ used to discretize the differentio-integral 
equations (\ref{evolf}) and  (\ref{evolfd}). In order to solve the two-point
boundary problem, we applied the iterative procedure defined by 
Braun~\cite{braun2}. Thus, in each iteration the evolution in rapidity 
of one of the field $f(y,k^2)$ or $\fda(y,k^2)$ was performed while the 
other field was set to its value obtained in the previous iteration. 
In the odd iterations, the field $f(y,k^2)$ was evolved from its initial 
value $f_A(k^2)$ from $y=0$ to $y=Y$ and the field $\fda(y,k^2)$ was 
kept fixed (in the first iteration $\fda(y,k^2)$ was set identically to zero).
In the even iterations $\fda(y,k^2)$ was evolved from $y=Y$ 
down to $y=0$ with the initial value $\fda_B(k^2)$ at $y=Y$. 
The iterations were continued until a fixed point of $f(y,k^2)$ 
and $\fda(y,k^2)$ was reached. Clearly, at the fixed point of the iterative 
procedure the fields $\{f(y,k^2),\fda(y,k^2)\}$ solve the 
system of Braun equations (\ref{evolf}) and (\ref{evolfd}) 
with boundary conditions (\ref{boundary}). 
The method was found to be stable and robust and no problems with 
convergence occurred of the kind reported in \cite{braun2}. 
A disadvantage of the iterative method is that it may be only used 
to find the solutions which represent the attractive fixed points of the
procedure. Unfortunately, we expect from the analysis of solutions of 
the RFT in zero transverse dimensions, that the solution to Braun equations 
is not unique at larger rapidities and there should exist multiple solutions.
Therefore it is probable that the iterative method finds only some of them. 
On the other hand, one hopes that the most relevant 
solutions that minimize the action may become an attractive fixed 
points of a reasonable iterative procedure. 
Some more arguments in favour of this scenario will be given 
based on the properties of the found solutions.

\subsection{Properties of solutions and spontaneous symmetry breaking}

In what follows we shall describe the properties of the solutions to the
Braun equations for rapidities of the scattering $Y=6,8,10,12$ and $Y=16$.
The pomeron field $\fda(y,k^2)$ will be often presented as a function of
the transformed rapidity $y'= Y-y$, so that the initial condition for 
$\fda$ is imposed at $y'=0$. In the figures we shall plot 
$f(y,k^2)/k^2$ and $\fda(y',k^2)/k^2$ unless we explicitly specify 
differently. Such a choice of variables
is preferred by their {\em solitonic} behaviour in the case of the
BK equation. For future reference, let us introduce the notation
$\fbk$ to represent the solution to the BK equation with the input
given by (\ref{input}). In the figures the label ``Input'' is used 
for  $f_A(k^2)/k^2$, where $f_A(k^2)$ defined by~(\ref{input}).

\begin{figure}[t]
\begin{center}
a) \psfig{file=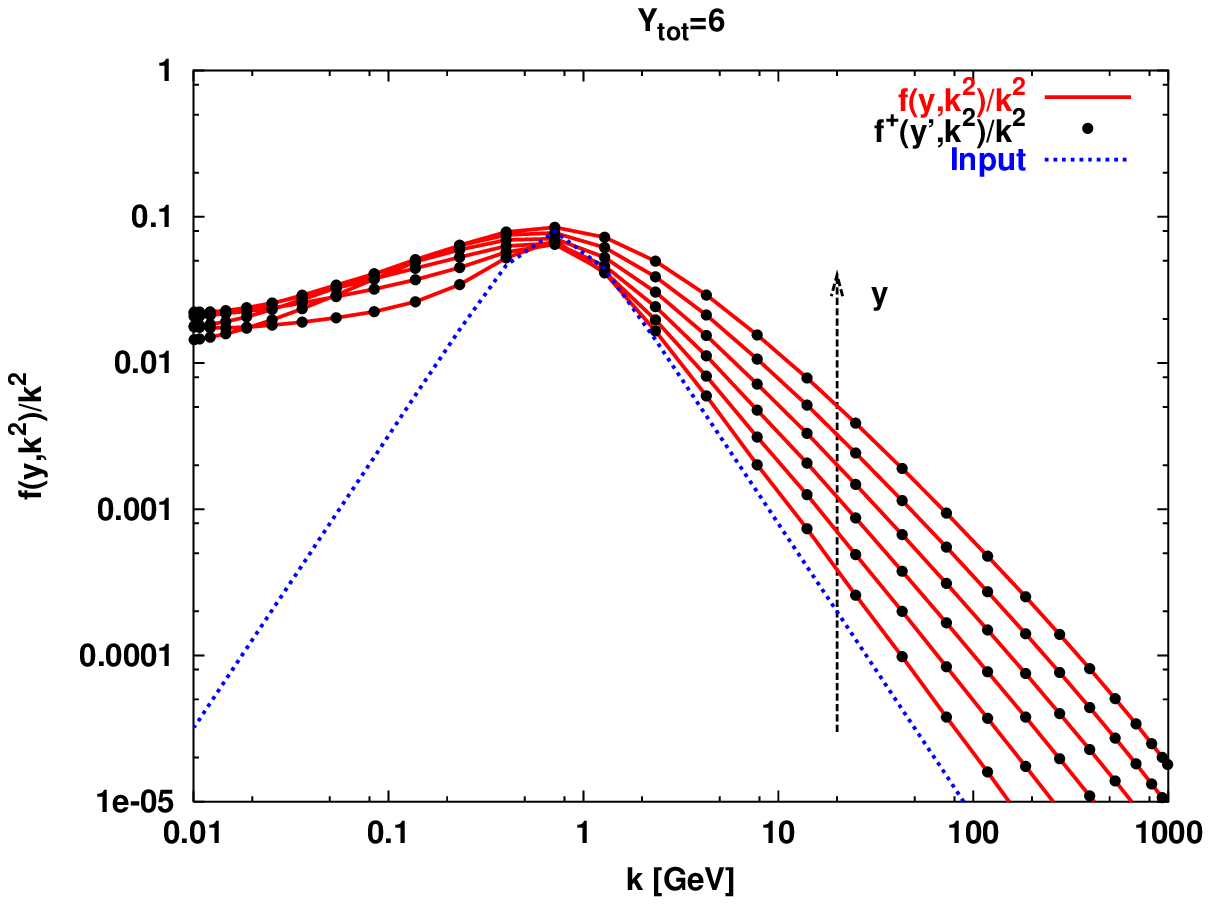,width=120mm} \\
b) \psfig{file=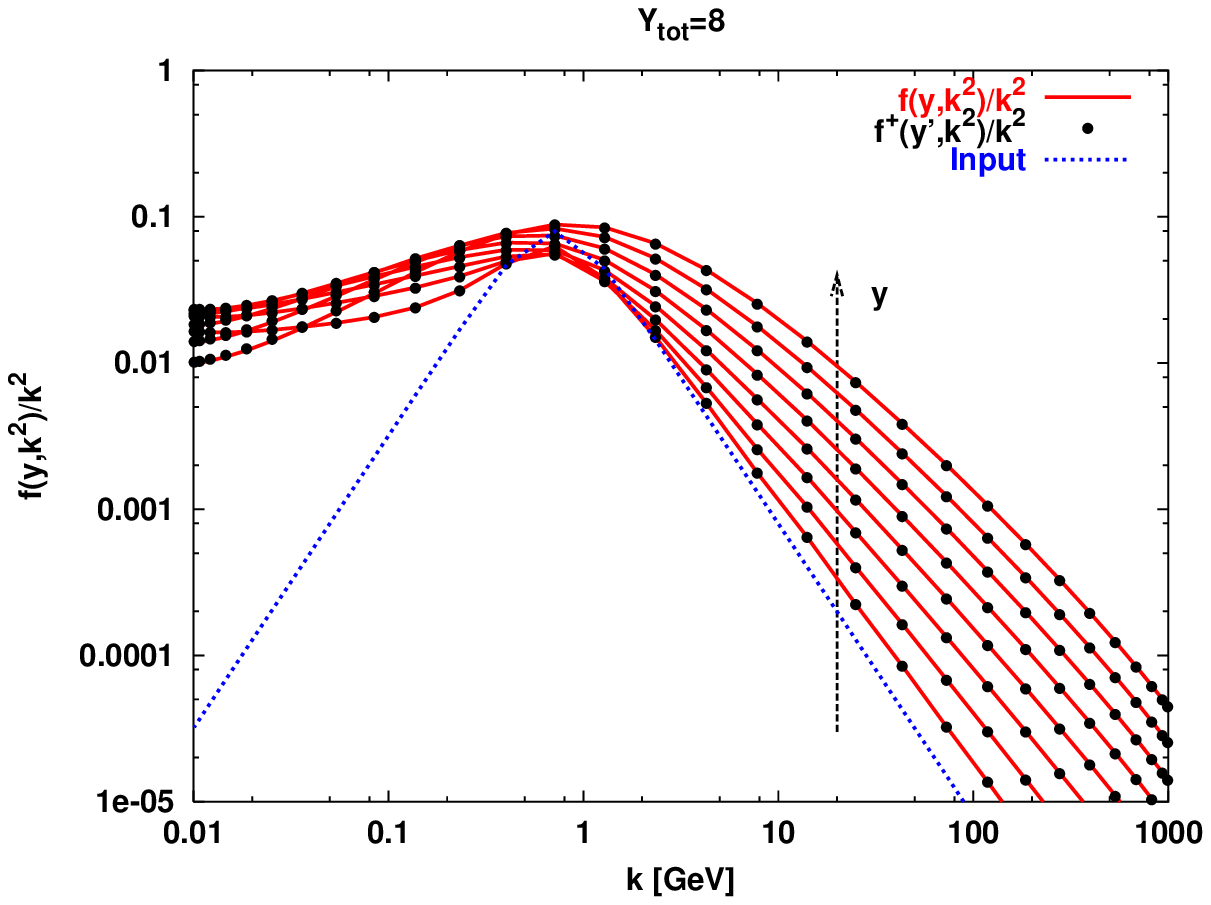,width=120mm} 
\end{center} 
\caption{\it Solutions of the Braun equations 
$f(y,k^2)/k^2=\fda(y',k^2)/k^2$ for a) $Y=6$ and b) $Y=8$.}   
\label{Res1}
\end{figure}

In \fig{Res1}a and \fig{Res1}b we show the solutions to the Braun
equations for $Y=6$ and $Y=8$ respectively. Solid lines 
denote $f(y,k^2)/k^2$ and $\fda(y',k^2)/k^2$ is shown with points. 
The curves are plotted for $y$ varying from zero to $Y$ in the steps 
of one. Clearly, in both cases the solutions are symmetric, 
$f(y,k^2)=\fda(Y-y,k^2)$. Anticipating \fig{Res5}a, we 
point out that the the solutions exhibit a similar behaviour to
solutions of the BK~equation at large gluon momenta $k^2$. 
At small momenta, below the saturation scale of the BK equation, 
$f(y,k^2)$ and $\fda(y',k^2)$ are much flatter than $\fbk$.

\begin{figure}[t]
\begin{center}
a)\psfig{file=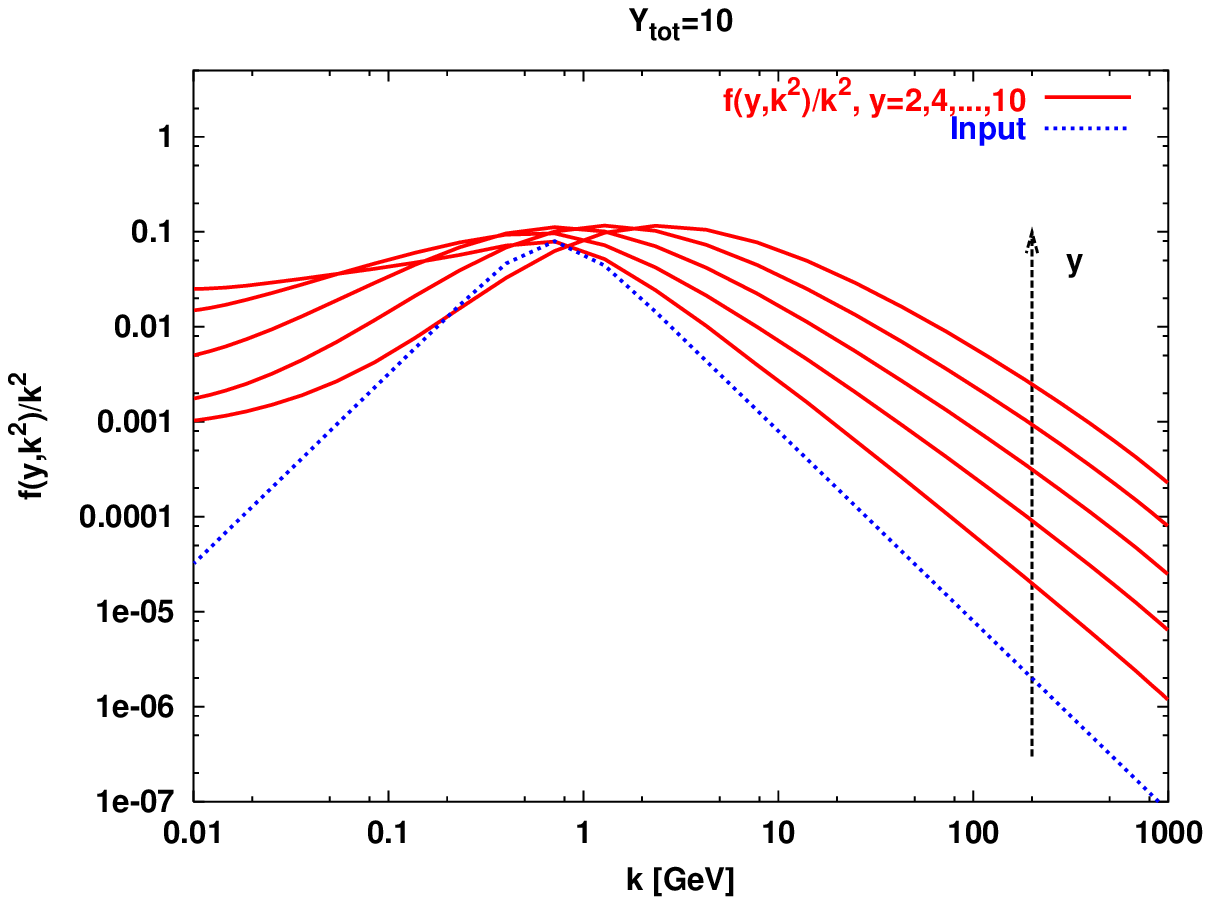,width=120mm} \\
b)\psfig{file=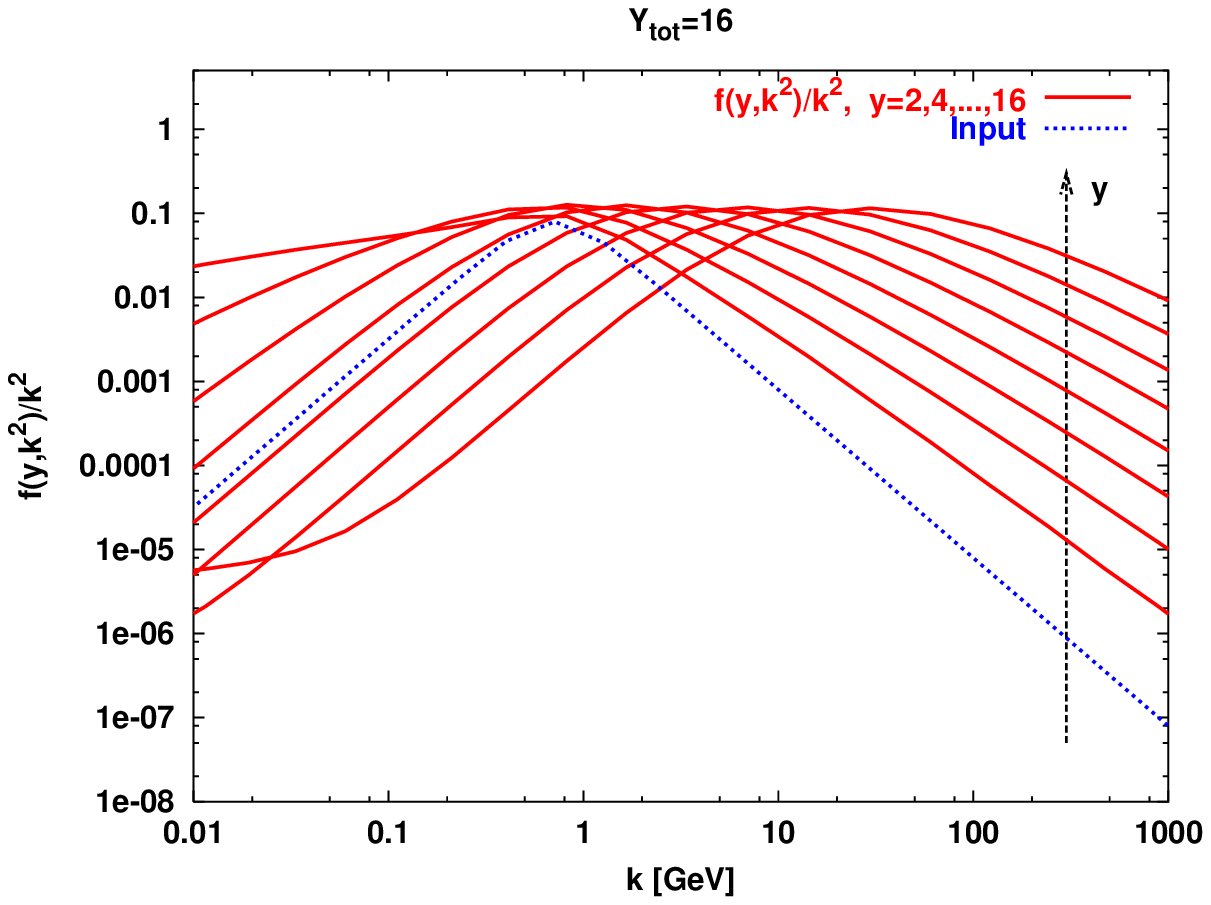,width=120mm} 
\end{center} 
\caption{\it 
Solutions of the Braun equations  $f(y,k^2)/k^2$ for a) $Y=10$ 
and b) $Y=16$.}
\label{Res2n}
\end{figure}

\begin{figure}[t]
\begin{center}
a)\psfig{file=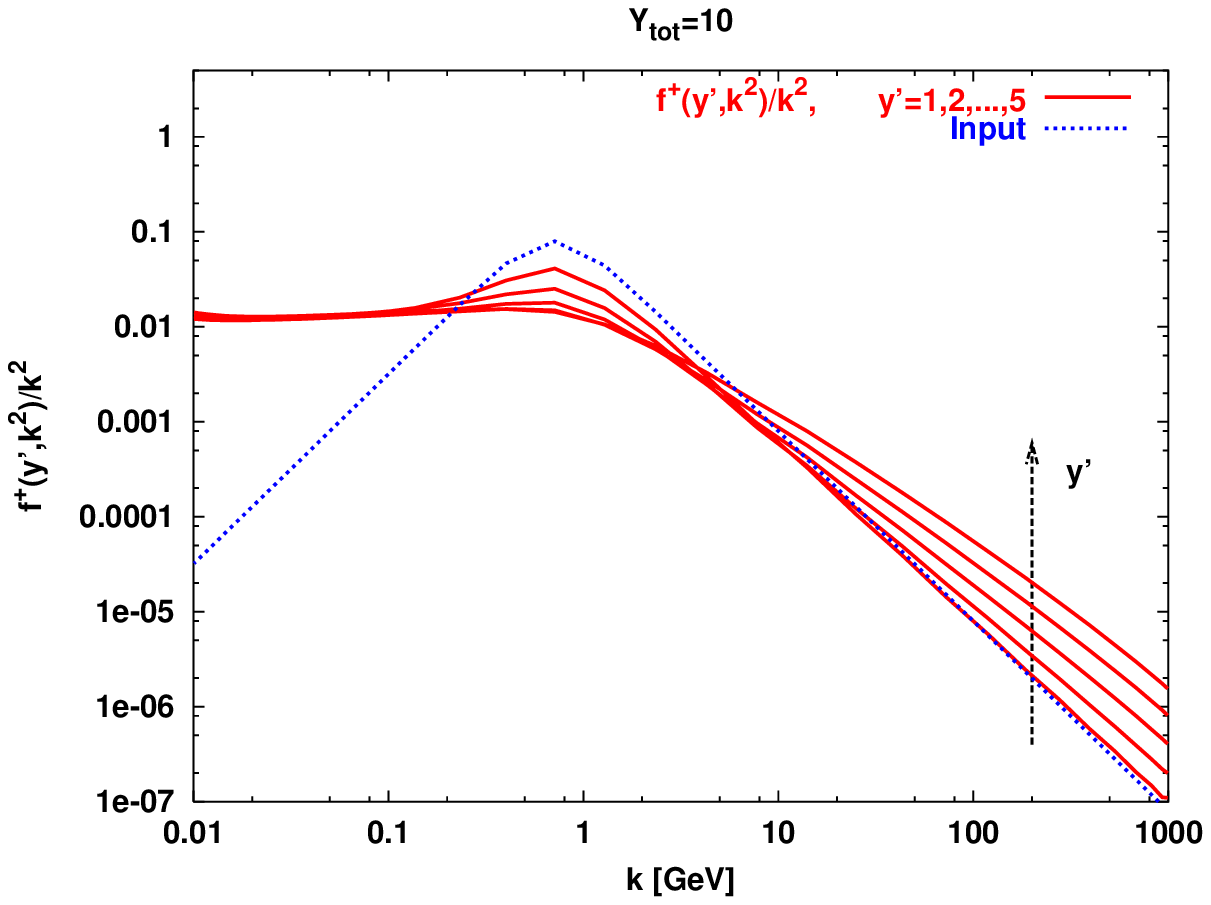,width=120mm} \\
b)\psfig{file=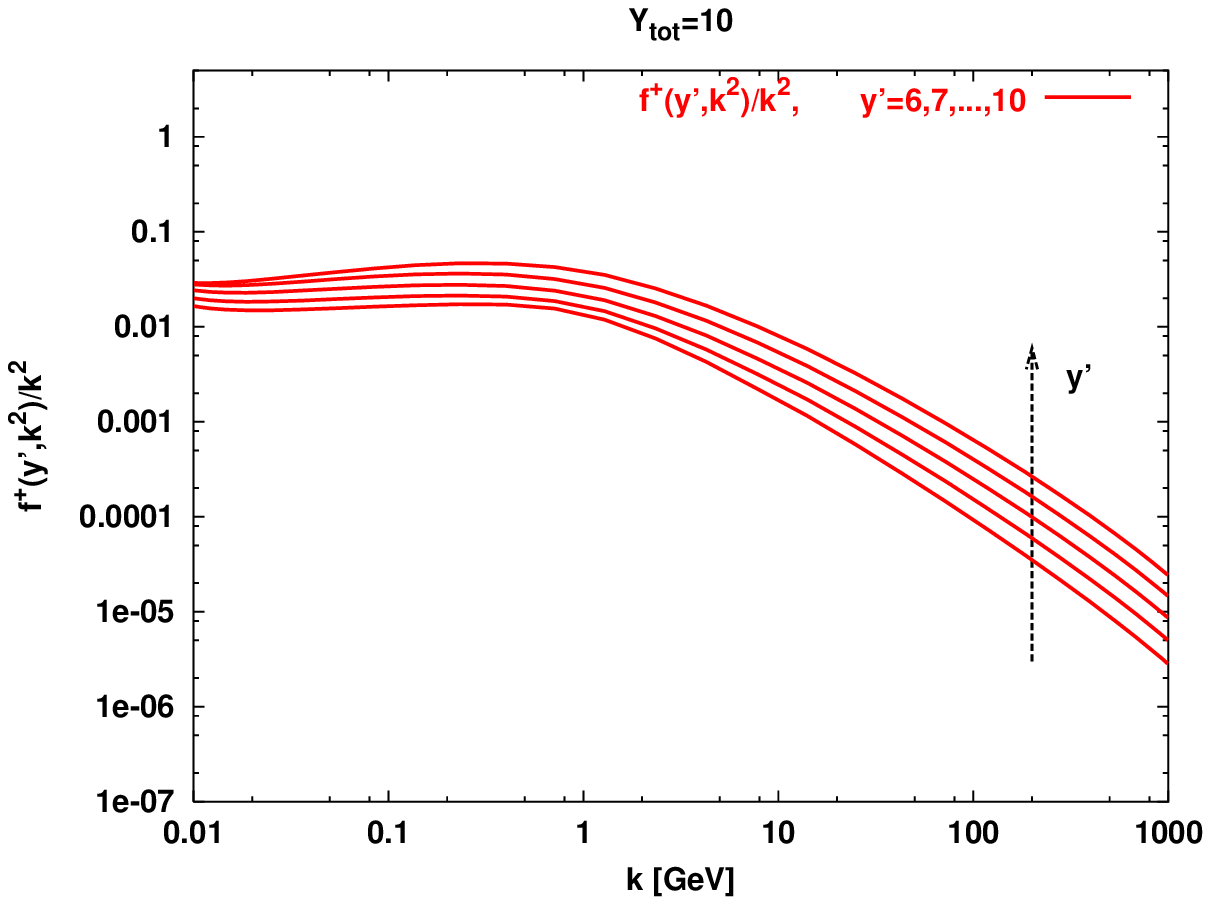,width=120mm} 
\end{center} 
\caption{\it Solutions of the Braun equations $\fda(y',k^2)/k^2$ for $Y=10$:
a) $y'=0,1,\ldots,5$;  
b) $y'=6,7,\ldots,10$.}   
\label{Res3n}
\end{figure}

\begin{figure}[t]
\begin{center}
a)\psfig{file=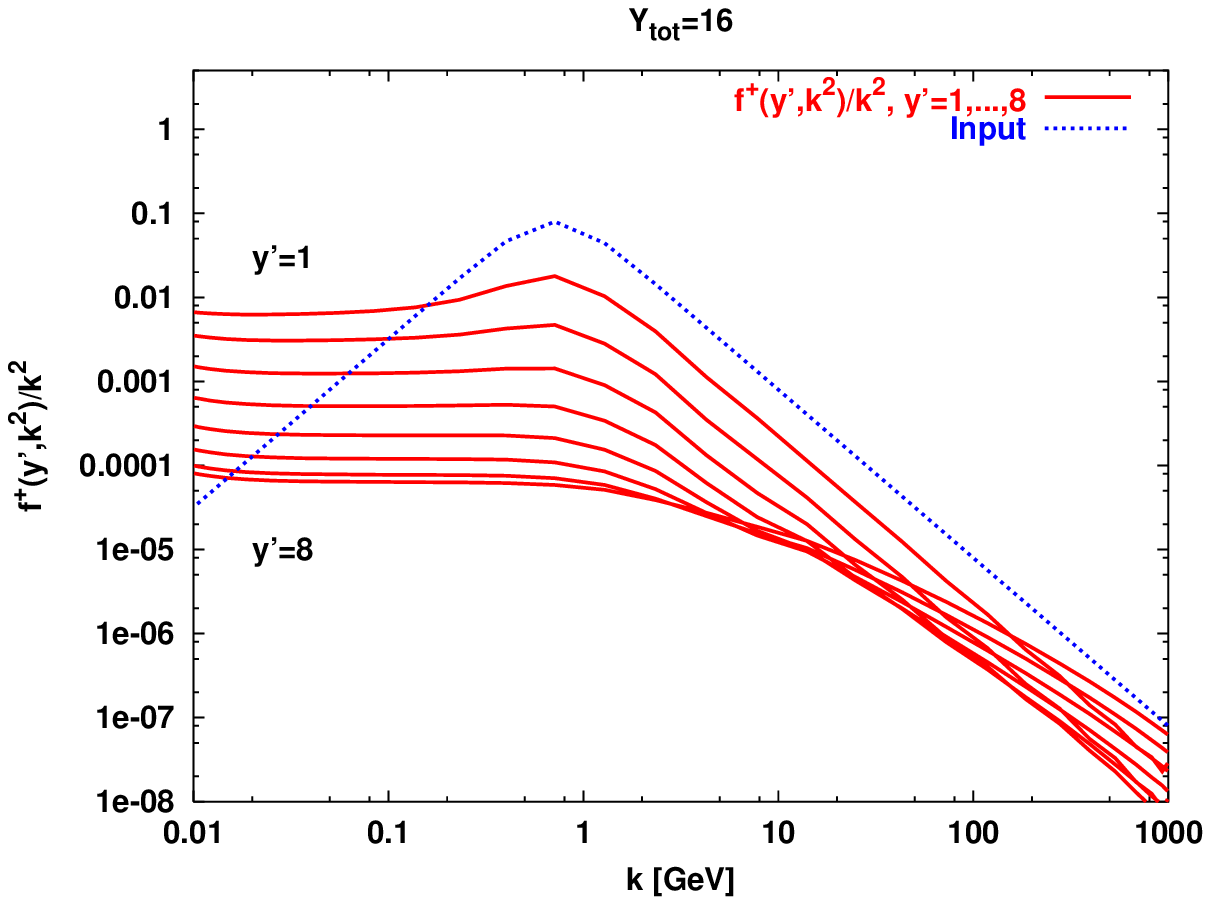,width=120mm} \\
b)\psfig{file=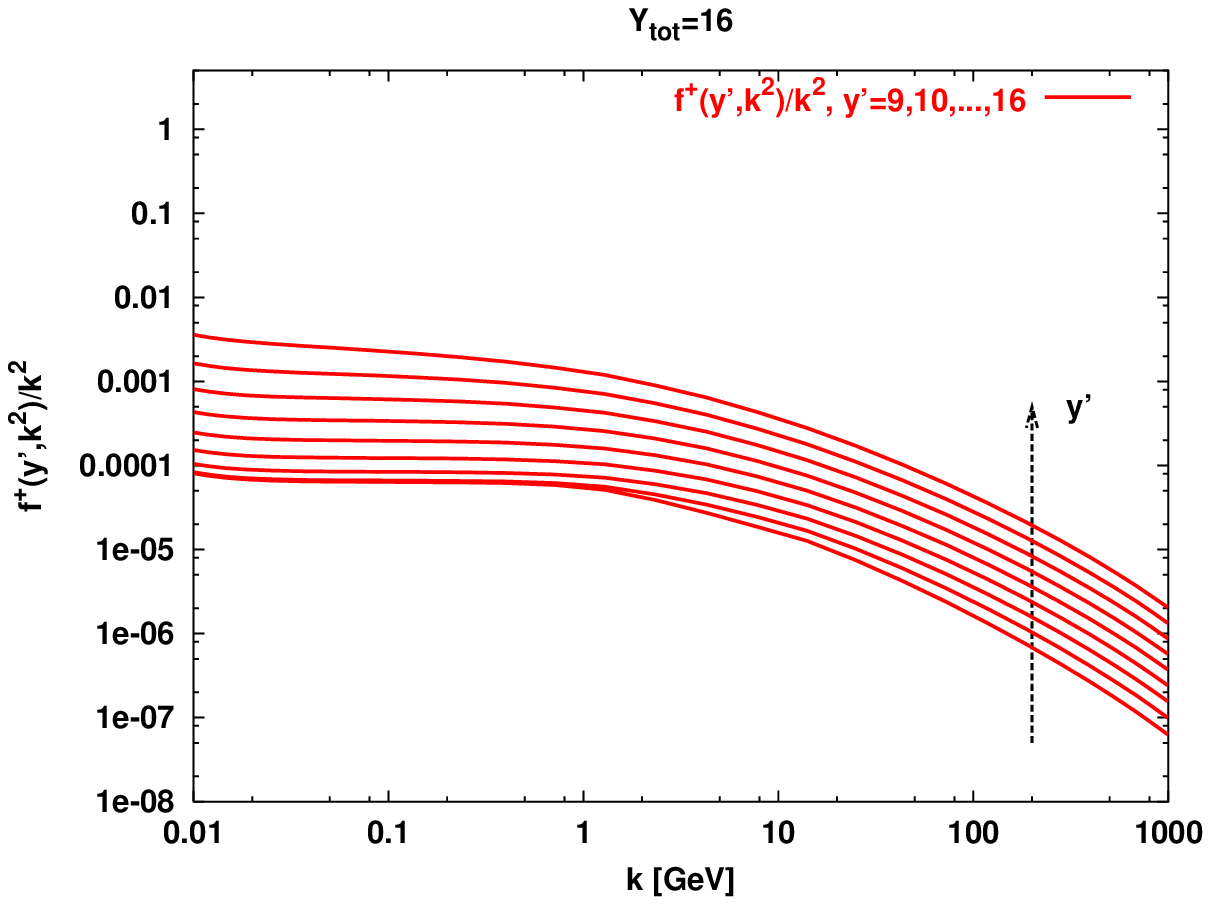,width=120mm} 
\end{center} 
\caption{\it Solutions of the Braun equations $\fda(y',k^2)/k^2$ for $Y=16$:
a) $y'=0,1,\ldots,8$;  
b) $y'=9,10,\ldots,16$.}   
\label{Res4n}
\end{figure}

We find that the symmetry between $f$ and $\fda$ breaks down at some 
critical rapidity $Y_c \simeq 9$. For $Y>Y_c$ only the
asymmetric solutions are found, for which $f(y,k^2) \neq \fda(Y-y,k^2)$,
compare Fig.~\ref{Res2n}a (Fig.~\ref{Res2n}b) 
and Fig.~\ref{Res3n} (Fig.~\ref{Res4n}) for $Y=10$ ($Y=16$).
Thus, the symmetry between the projectile and the target is spontaneously
broken for individual classical solutions, in close analogy with the
phenomenon appearing in RFT-0, described in detail in  Sec.~\ref{sec:ssb0}. 
The asymmetry between $f(y,k^2)$ and $\fda(Y-y,k^2)$ vanishes at $Y=Y_c$, 
so the asymmetric solutions connect smoothly to the symmetric one 
at $Y=Y_c$ and the asymmetry  builds up gradually with increasing $Y$.
Certainly, the numeric value of the critical rapidity~$Y_c$ 
is not universal, it depends on the boundary conditions and on the
value of $\alpha_s$. It is important to note, that for each asymmetric
solution $\{f,\fda\}$ there exists a complementary solution 
$\{f',{f^{\dagger}}'\}$, 
such that $f'(y,k^2) = \fda(Y-y,k^2)$ and ${f^{\dagger}}'(Y-y,k^2) = f(y,k^2)$,
reflecting the symmetry between the projectile and the target encoded
in the action and the symmetric boundary conditions.
Knowing that, in the further analysis of the solutions we choose 
arbitrarily that $f(y,k^2)$ is the larger field and $\fda(y',k^2)$ 
is the smaller one.

For $Y>Y_c$, the general features of the larger field $f$ are the following.
At $Y\simeq Y_c$ the solution is similar to the symmetric solutions 
found for $Y<Y_c$. With increasing $Y$ a pattern appears of a 
traveling wave, that is formation of a peak of $f(y,k^2)/k^2$ traveling 
towards larger values of $\log(k^2)$ with increasing rapidity with 
only small changes of the shape, see Fig.~\ref{Res2n}a and Fig.~\ref{Res2n}b.
Recall, that it is behaviour characteristic for the BK 
equation~\cite{traveling}. The similarity of the solution to the BK 
solution will be investigated in more detail in Sec.~\ref{sec:bkfan}.

\begin{figure}[t]
\begin{center}
a)\psfig{file=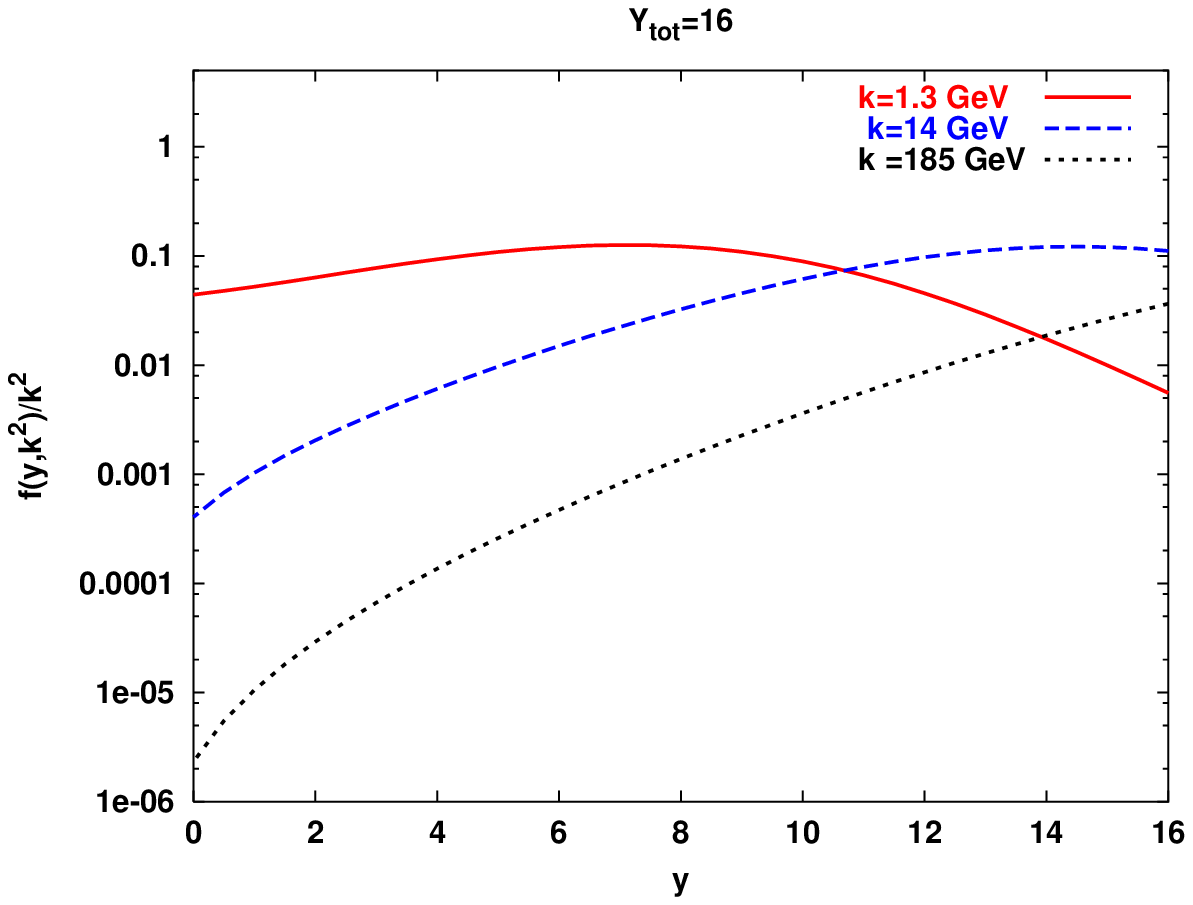,width=110mm} \\
b)\psfig{file=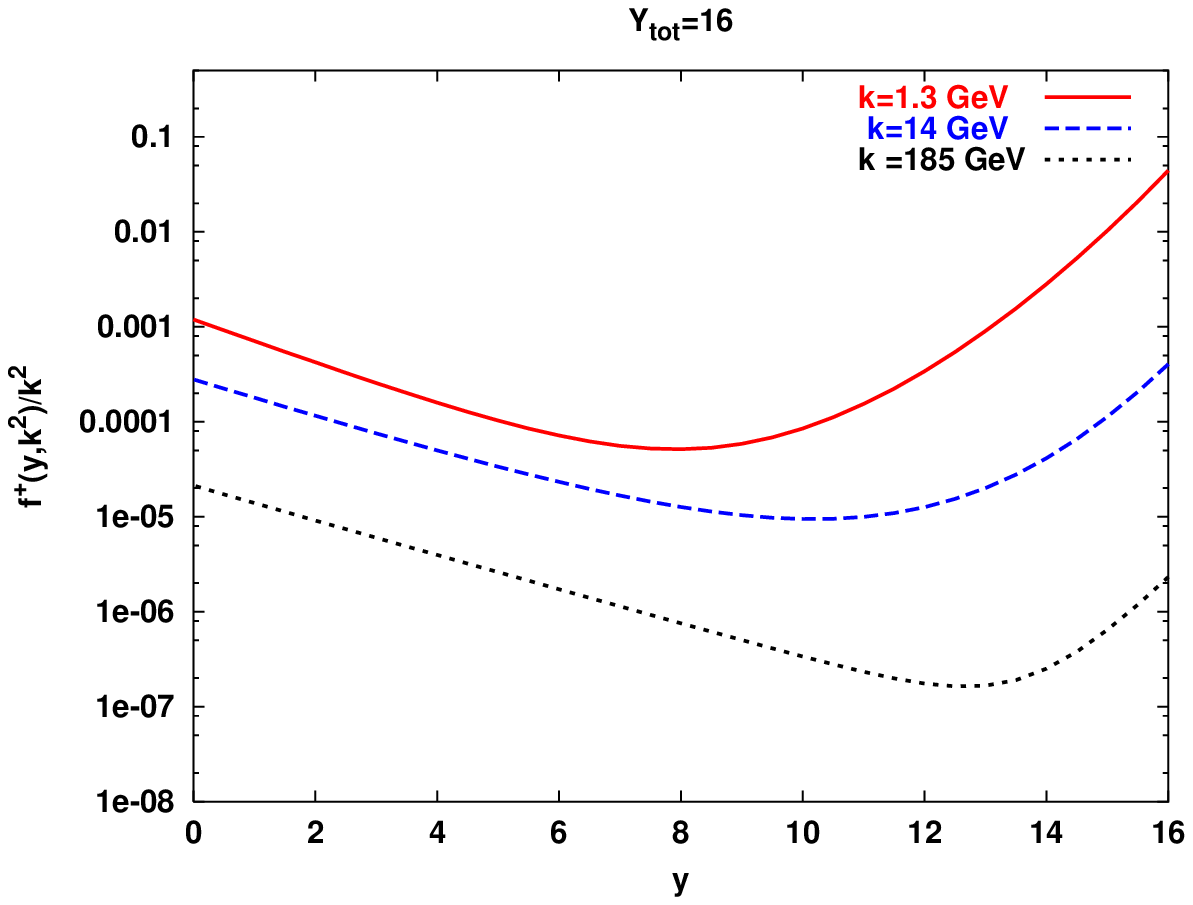,width=110mm} 
\end{center} 
\caption{\it Solutions of the Braun equations for $Y=16$ plotted 
as a function of rapidity $y$ for $k=1.3$~GeV (solid line), 
$k=14$~GeV (dashed line) and $k=185$~GeV (dotted line):
a) $f(y,k^2)/k^2$ and  b) $\fda(y,k^2)/k^2$.}   
\label{Res4.5}
\end{figure}

The smaller field $\fda(y,k^2)$ evolves differently, see Fig.~\ref{Res3n} and
Fig.~\ref{Res4n}. At $Y=Y_c$ it matches $f(Y-y,k^2)$ and for the increasing 
$Y$ it experiences a significant overall suppression, stronger at larger $Y$. 
For instance, for $y \simeq Y/2$ the maximal value of  $\fda(y,k^2)/k^2$ 
is about an order of magnitude smaller than the maximal value of  
$f(y,k^2)/k^2$ at $Y=10$ (compare Fig.~\ref{Res2n}a and Fig.~\ref{Res3n})
and about three orders of magnitude smaller at $Y=16$ 
(see Fig.~\ref{Res2n}b and Fig.~\ref{Res4n}).
Thus, we conjecture that in the limiting case of very large
total rapidity $Y$, $\fda$ may is arbitrarily small except of the 
rapidities $y\simeq Y$ where the source term for $\fda$ is still 
important. In this context, it is instructive to study the $y$-dependence
of $\fda(y,k^2)/k^2$ at fixed~$k$ and compare it to $f(y,k^2)/k^2$. This
comparison may be performed using Fig.~\ref{Res4.5}. It turns out, 
that there appear two distinct regimes of evolution of  $\fda(y,k^2)/k^2$ with 
rapidity (at fixed momentum). Thus, if the field $f(y,k^2)/k^2$ is strong,
the smaller field $\fda(y',k^2)/k^2$ is exponentially suppressed with 
increasing $y'$, $\fda(y',k^2)/k^2 \sim \exp(-\beta_1 y')$
with $\beta_1 \sim 1$, crudely. This is the region
where the absorption of $\fda$ by $f$ drives the evolution of $\fda$. 
Then, when $y$ is sufficiently small and the field  $f(y,k^2)/k^2$ is 
weaker, the absorption becomes less relevant and $\fda(y',k^2)/k^2$ grows
exponentially with increasing $y'$,  $\fda(y',k^2)/k^2 \sim \exp(\beta_2 y')$
with the exponent $\beta_2 \simeq 0.4$ (with our choice of parameters)  
a value somewhat smaller than the BFKL intercept 
$\omega_0 = 4\bar\alpha_s\log(2) \simeq 0.53$.
Note, that the characteristic rapidity $y$, at which 
the transition occurs from the strong absorption regime to the BFKL 
driven growth regime, depends on~$k$. This is natural, as the field
$f(y,k^2)/k^2$ becomes strong at larger values of~$y$ for larger~$k$.

The shape of $\fda(y',k^2)/k^2$ in $k^2$ exhibits some interesting features 
too, see Fig.~\ref{Res3n} and Fig.~\ref{Res4n}.
At small values of~$k$, $\fda(y',k^2)/k^2$ tends to a flat function.
This should be compared with the case of the BK where $\fbk/k^2 \sim k^2$
at small~$k$. On the other hand, at large $k^2$ the decrease of 
$\fda(y',k^2)/k^2$ with increasing $k^2$ is slower than the decrease
of $\fbk/k^2$. Thus, the overall picture is that $\fda(y',k^2)/k^2$
is much flatter than $\fbk/k^2$. 
As rather surprising comes an observation that effects 
of non-linear interactions in $\fda(y',k^2)/k^2$ extend to very large 
values of momenta, causing a strong suppression of $\fda(y',k^2)/k^2$ 
for all momenta up to  $k=10^3$~GeV, the value larger than the saturation 
scale generated by the large field $f(y,k^2)$, 
see for example Fig.~\ref{Res4n}.
In fact, we checked that the suppression of $\fda(y',k^2)/k^2$ in comparison
to $\fbk/k^2$ is strong even at $k=10^5$~GeV (not shown).

The explanation of this phenomenon is the following. At large 
values of $k^2$ the input function $\fda_B(k^2)$ was assumed to tend to a 
constant, in other words the anomalous dimension vanished for the input. 
For the BFKL or the BK system the rapidity evolution generates an anomalous 
dimension of $\gamma_0 \simeq 0.3-0.5$, strongly enhancing $f(y,k^2)$
for large $k^2$ and $y$. For $\fda(y',k^2)$, however, the evolution and 
BFKL diffusion are almost completely blocked by large absorptive corrections
coming from the interaction of $\fda(y',k^2)$ with the large field  
$f(y,k^2)$. Recall, that the input for $\fda(y',k^2)$ resides in $y=Y$,
where the value of the field $f(y,k^2)$ is the largest and so is the related
saturation scale. Therefore, before any BFKL diffusion or enhancement 
of $\fda(y',k^2)$ becomes possible (that is at sufficiently small $y'$)
strong suppression of $\fda(y',k^2)$ occurs and the population of the
large momenta region is initiated from a drastically reduced $\fda(y',k^2)$.

\begin{figure}[t]
\begin{center}
a)\psfig{file=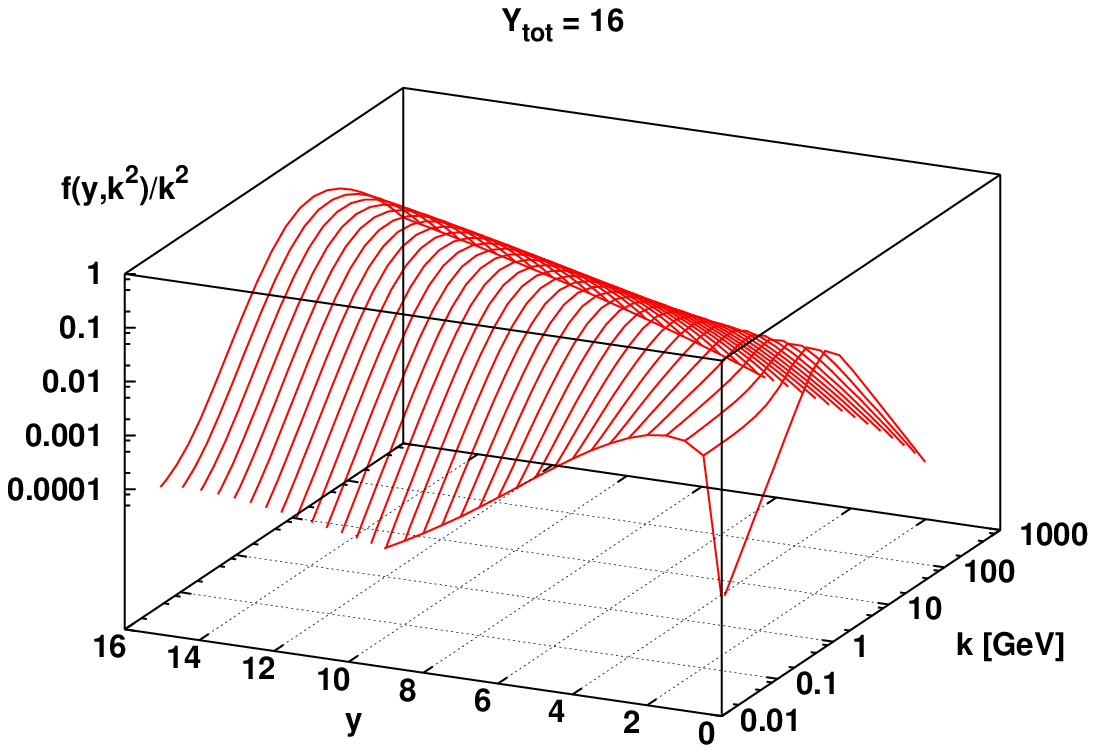,width=160mm} \\
b)\psfig{file=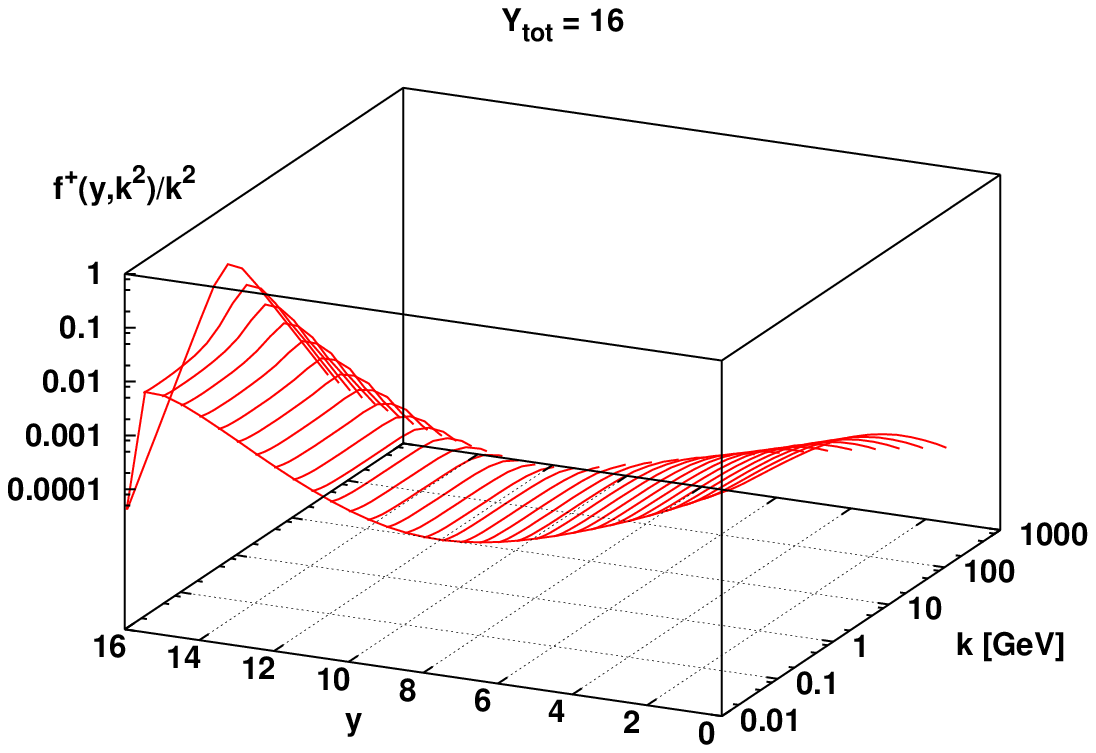,width=160mm} 
\end{center} 
\caption{\it Solutions of the Braun equations for $Y=16$ plotted 
as a function of rapidity $y$ and $k$: 
a)~$f(y,k^2)/k^2$ and  b)~$\fda(y,k^2)/k^2$.}   
\label{Res3d}
\end{figure}

In order to provide a more synthetic picture of the behaviour 
of $f(y,k^2)$ and $\fda(y,k^2)$ we illustrate the case of $Y=16$
with three dimensional plots of the solutions shown in Fig.~\ref{Res3d}.
Note, that we plot in this figure $\fda(y,k^2)$ instead of $\fda(y',k^2)$. 
Thus, the input for $f$ appears at $y=0$ in Fig.~\ref{Res3d}a and the
input of $\fda$ is plotted for $y=16$ in  Fig.~\ref{Res3d}b.

\subsection{BK fan dominance}
\label{sec:bkfan} 

\begin{figure}[t]
\begin{center}
a)\psfig{file=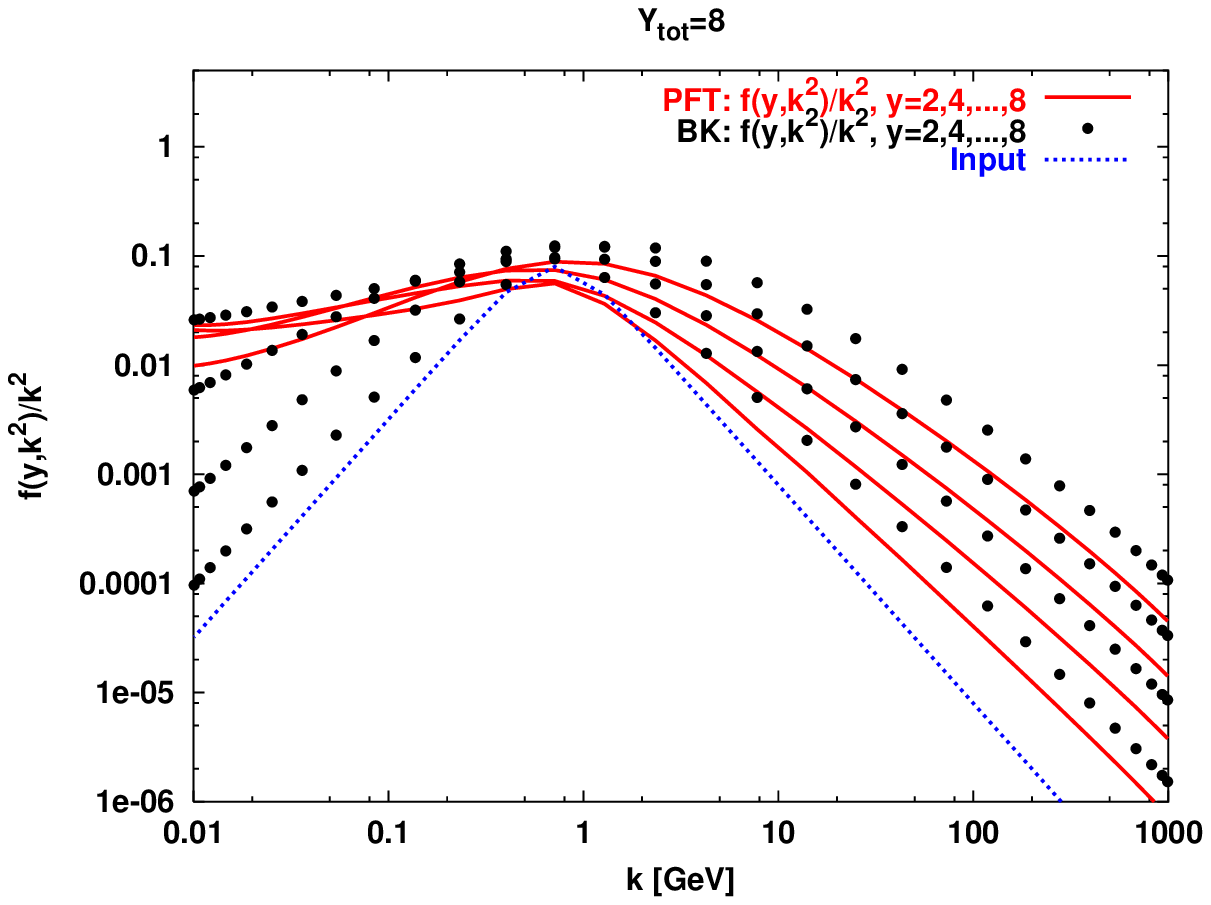,width=120mm} \\
b)\psfig{file=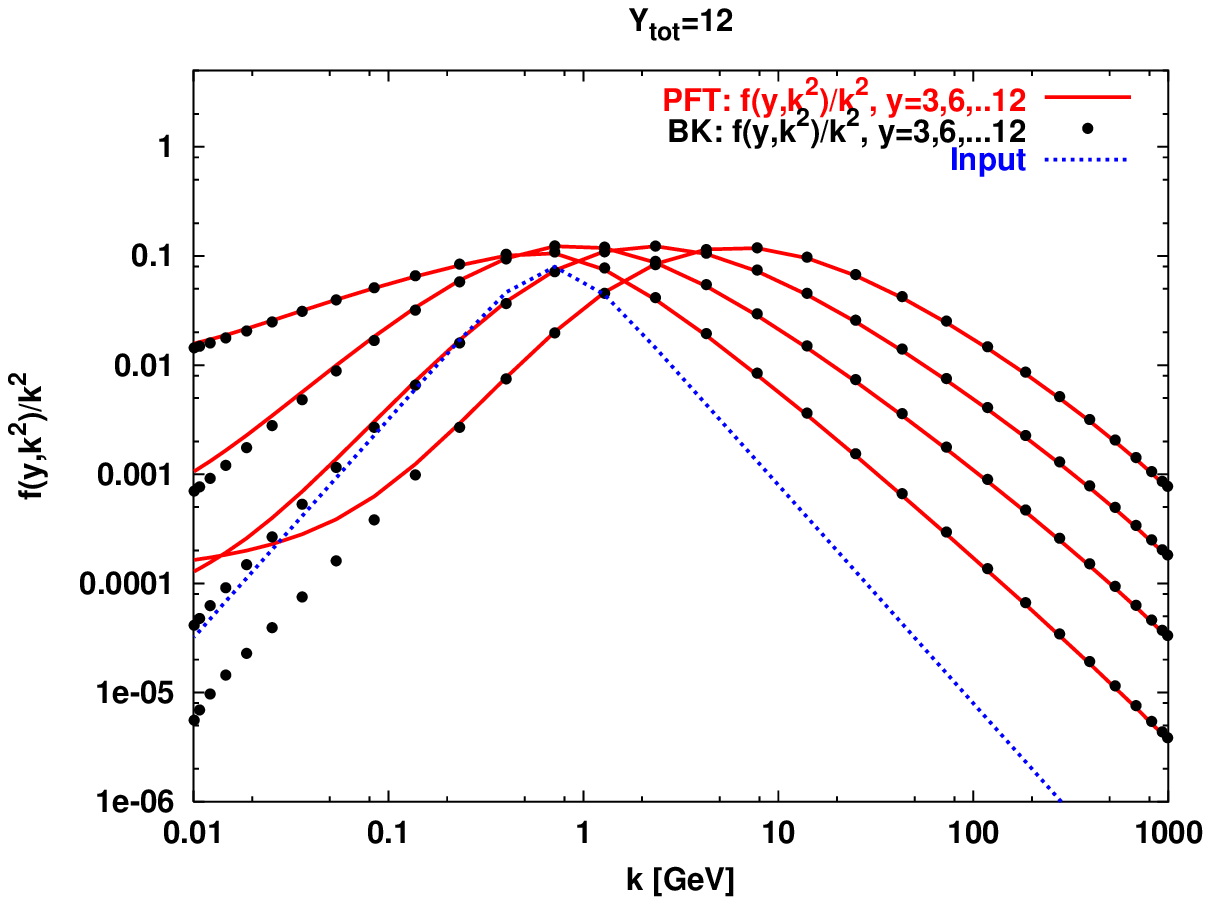,width=120mm} \\
\end{center} 
\caption{\it Comparison of the larger solution $f(y,k^2)/k^2$ 
of the Braun equations (solid line) to the solution of the 
BK equation (points) for a) $Y=8$ and b) $Y=12$.}   
\label{Res5}
\end{figure}

\begin{figure}[t]
\begin{center}
a)\psfig{file=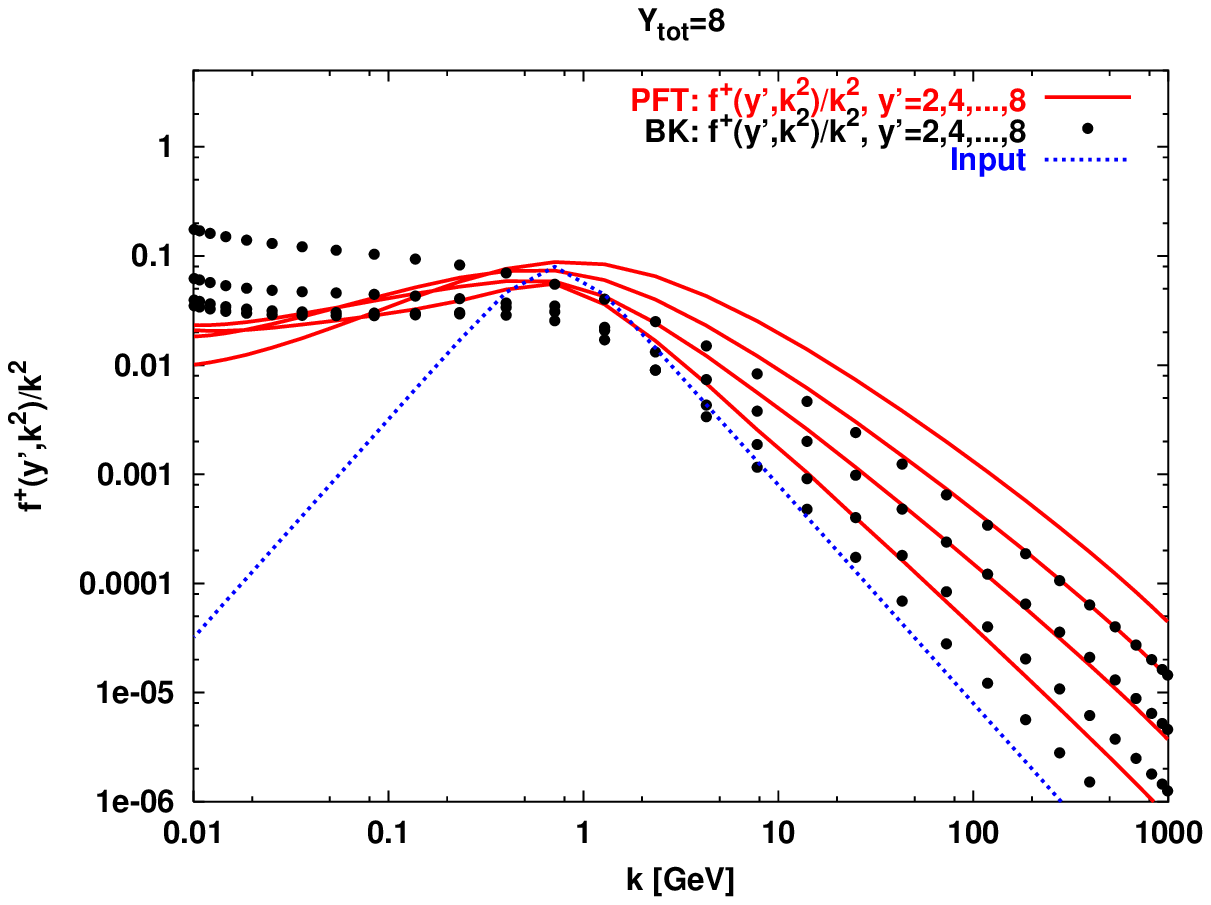,width=120mm} \\
b)\psfig{file=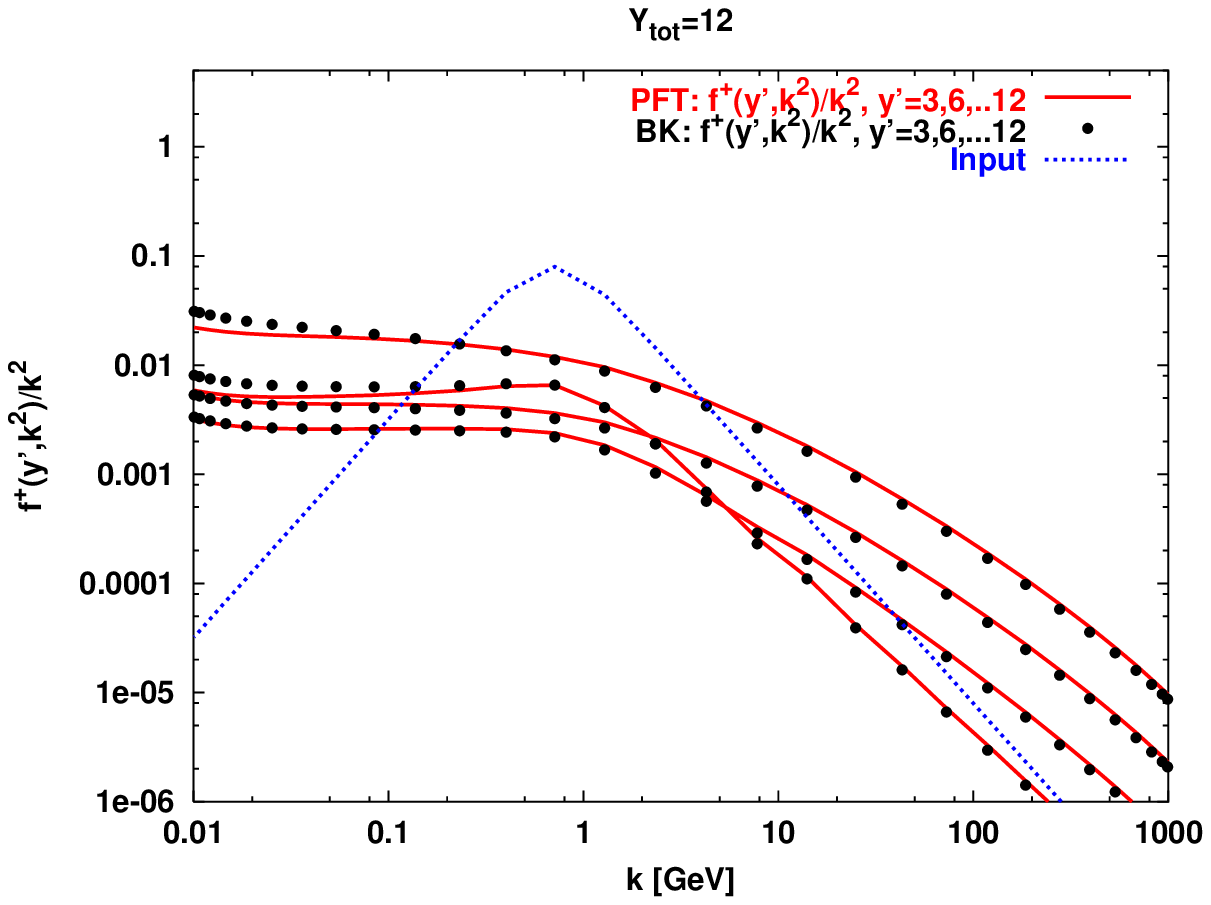,width=120mm} \\
\end{center} 
\caption{\it Comparison of the smaller solution $\fda(y',k^2)/k^2$ 
of the Braun equations (solid line) to the ``smaller solution'' 
$\fbkd/k^2$ of the BK equation (points) for a) $Y=8$ and b) $Y=12$.}   
\label{Res6}
\end{figure}

\begin{figure}[t]
\begin{center}
a) \psfig{file=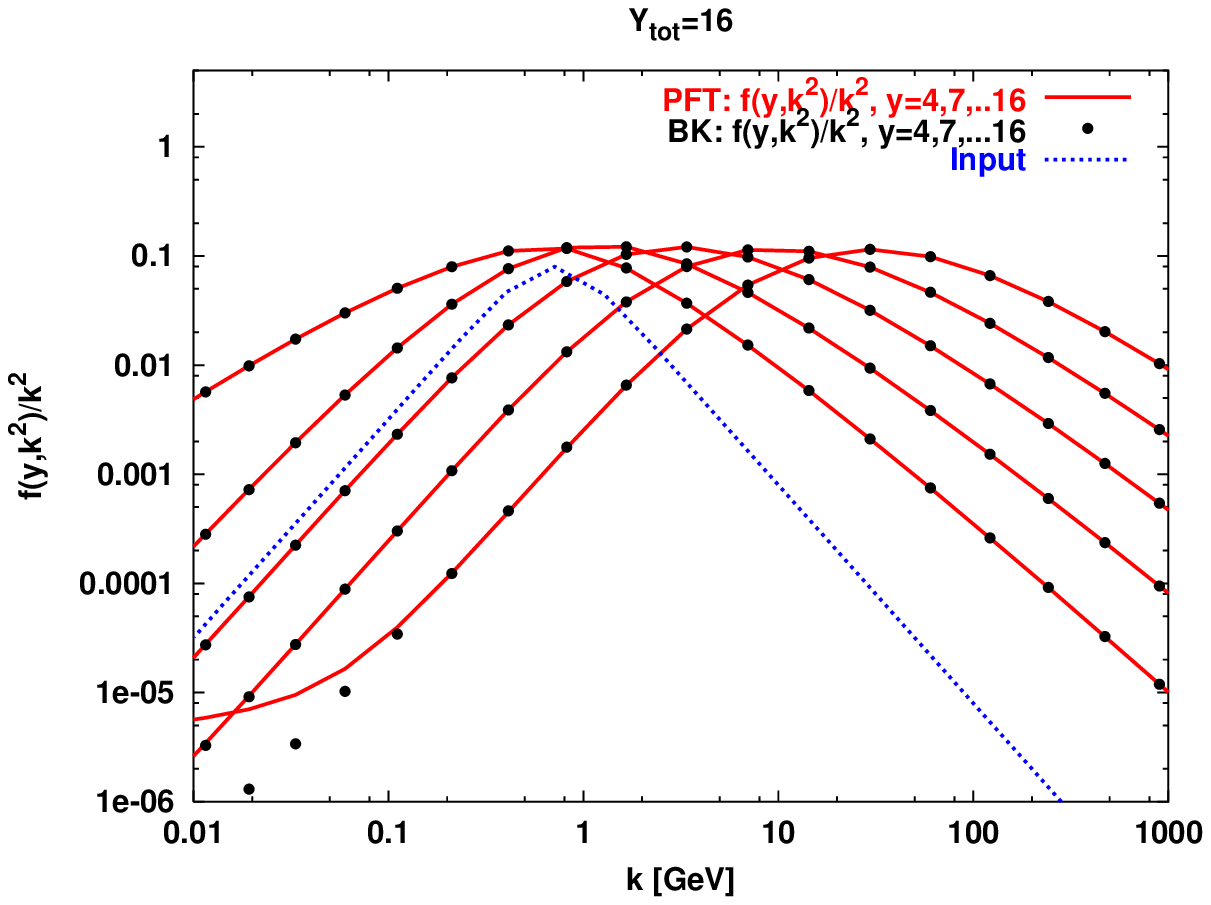,width=120mm} \\ 
b)\psfig{file=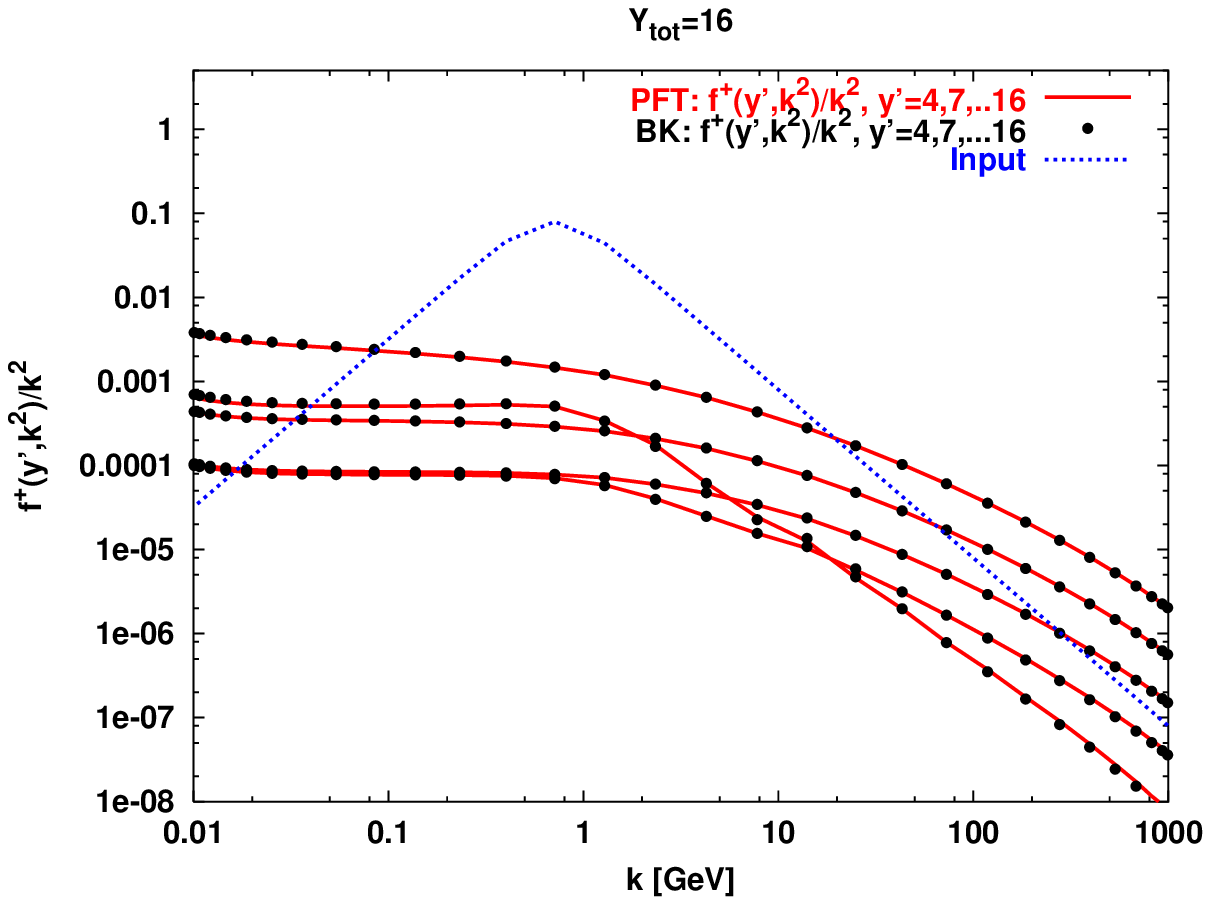,width=120mm} \\
\end{center} 
\caption{\it Comparison of the solution of the Braun equations 
(solid line) to the solution of the BK equation (points) for $Y=16$:
a) the larger solution $f(y,k^2)/k^2$ and b) the smaller solution 
$\fda(y',k^2)/k^2$.
}   
\label{Res7}
\end{figure}

We have already related briefly the larger component  $f(y,k^2)$
of the solution to the Braun equation to the solution of the 
Balitsky-Kovchegov equation $\fbk$. A more detailed comparison is 
performed for  $f(y,k^2)$ in \fig{Res5} for $Y=8$ and $Y=12$, and in 
\fig{Res7}a for $Y=16$. 
For $Y=8$ where the solution is still symmetric, the difference between
$f(y,k^2)$ (lines) and $\fbk$ (points) is quite large and low $k^2$ and
visible at large $k^2$, see \fig{Res5}a. The difference is significantly
reduced at $Y=12$, as clearly seen in \fig{Res5}b. Here, $f(y,k^2)$ and 
$\fbk$ almost exactly coincide except of some deviations for very small 
$k<0.1$~GeV and $y > Y/2$. The overlap between  $f(y,k^2)$ and $\fbk$  
is further improved at $Y=16$. Evolution of the smaller component 
$\fbkd$ in the BK limit may be also performed by solving the system 
\eq{evolf} and \eq{evolfd} with the terms neglected that were 
generated by the triple pomeron vertex corresponding to the pomeron 
splitting (the contribution to the action of ${\cal L}_3 ^\dagger$). 
The comparison of $\fda(y',k^2)$ and $\fbkd$ is given for 
$Y=8$, $Y=12$ in \fig{Res6} and for $Y=16$ in \fig{Res7}b.
In this case, the two different kinds of solutions coincide even better
than the large components $f(y,k^2)$ and $\fbk$.

Recall, that we also observed the similarity between the asymmetrical 
classical solutions of RFT-0 and the solution to the ``fan equation'',
the counterpart of the BK~equation in zero transverse dimensions. 
Thus, the ``fan dominance'' at $Y \gg Y_c$ seems to be a generic feature 
of the interacting pomeron system. Even more can be said -- the dependence
of the QCD pomeron fields on the momentum might even enhance the convergence
to the fan dominated system, compare \fig{RFT6} and \fig{Res7}.
It happens probably because the deviations from the fan behavior appear at
rapidities $y \to Y$, close to the source of the smaller field 
(which we chose to be $\fda(y,k^2)$ for QCD pomerons and $p(y)$ for RFT-0) 
where the smaller field is not yet strongly suppressed by the evolution. 
In QCD, however, the input is localized at rather small values of 
gluon momenta~$k$, whereas the larger field, $f(y,k^2)$, is concentrated
around the saturation scale $Q_s(y)$ which is large for $y\to Y$. Therefore, 
the relatively large $\fda(y',k^2)$ in this rapidity domain
affects only the tail of low momenta in $f(y,k^2)$, with little relevance 
for the dynamics of the system. Possible implications of the 
``fan dominance'' are discussed in the Sec.\ \ref{sec:disc}.

As the last point, let us comment shortly on the issue of multiple solutions
to the Braun equations out of which only some can be found by the 
iterative solving procedure. Recall, that in the semi-classical 
approximation of the system moving in the Euclidean time, the most relevant
are the trajectories with the lowest value of the action. We have no proof
that the asymmetric classical trajectories that were found in this paper 
fulfill this requirement. Numerous similarities of the patterns of 
solutions obtained in the interacting QCD Pomeron Field Theory and RFT-0 
are, fortunately, reassuring. In both cases there exists a symmetric solution
at low rapidity and two asymmetric solutions at $Y>Y_c$. In both theories 
the ``fan dominance'' phenomenon was found. Therefore, one may conjecture
that the asymmetric solutions of the Braun equations, indeed, are the 
classical trajectories with the lowest action, in analogy to the explicit 
result obtained in RFT-0.

\subsection{Summary of the results}

Let us summarize the presentation of results with a 
recapitulation of the most important observations:
\begin{enumerate}

\item Solutions $\{f,\fda\}$ of the Braun equations with symmetric 
boundary conditions split into two different types: the symmetric 
solution $f(y,k^2)=\fda(Y-y,k^2)$ that dominates below the critical 
rapidity $Y_c$ and a pair of the asymmetric solutions, found for $Y>Y_c$.

\item The asymmetric solutions exhibit the feature of ``fan dominance''
which becomes more accurate with increasing $Y$. Due to the smallness of one 
of the field the system evolves as if one of the triple pomeron vertices 
(describing splitting or merging) was absent. The larger field is close 
to the solution of the BK equation. 

\item Unitarity corrections for the smaller field $\fda(y',k^2)$ 
are very pronounced at $Y\gg Y_c$ leading to very strong damping 
(even 2-3 orders of magnitude at $Y=16$, and increasing with~$Y$) 
of the smaller field and flattening of the shape of $\fda(y',k^2)/k^2$. 
We find that $\fda(y',k^2) \sim k^2$ for $k^2 < Q^2 _s(y)$, where  
$Q_s(y)$ is the saturation scale generated by the
larger field $f(y,k^2)$.

\item The symmetric solution below the critical rapidity is significantly 
flatter at low momenta than the BK~solution. At large momenta the decrease 
of the symmetric solution is power-like, with an exponent close to that of 
BK, but the solution to the Braun system is somewhat smaller the the 
BK~solution from the same input.

\end{enumerate}

\section{Discussion}
\label{sec:disc}
The breaking of the projectile-target symmetry which
we have found above the critical rapidity is rather surprising and it calls
for an explanation and interpretation. To our understanding the mechanism 
of this breaking is the following. Suppose that we have a symmetric 
situation in the system of the two evolving pomeron fields $\{f,\fda\}$.
If the fields are small and therefore weakly interacting then the
interaction is only a small perturbation and the symmetry of the 
action and of the initial conditions should be reflected in the solution.
This is, indeed, the case for the Braun equations with the total rapidity~$Y$
smaller than the critical value $Y_c$. Let us consider now a symmetric 
system of pomeron fields when the fields are already strong due to
their rapidity evolution. Then, the fields absorb intensively each other.
The combination of the multiplication of the fields and the strong mutual
absorption is a potential source of instability. 
Namely, if we perturb the symmetric system of fields in this regime by, say, 
small increase of the value of the field~$f$ then the absorption 
of $\fda$ by interaction with $f$ will be also increased. 
Thus, after this perturbation $\fda$ should become smaller. 
This results, however, in a smaller absorption of the field $f$, 
leading to yet higher values of $f$, so that the instability will 
self-amplify generating finally an asymmetric configuration. 
Of course, whether this scenario is realized, depends on the particular form 
of the action. From our numerical results we conclude that this is, indeed, 
the case for the interacting pomeron fields above the critical rapidity
in the classical approximation. It is curious that in the earlier study of 
the Braun equations the symmetry breaking was not found~\cite{braun2}. 
Instead, there was reported an instability of the iterative procedure at 
critical values of rapidity, depending on the input. It was interpreted 
as a possible indication of a phase transition. We speculate that those 
instabilities might be, in fact, signs of emergence of asymmetrical solutions.

The key question arises what happens with the spontaneous symmetry
breaking (SSB) when the quantum effects are considered. Strictly speaking, for
finite systems SSB does not occur at the quantum level. 
The ground state of the finite system in which the symmetry is spontaneously
broken at the classical level is a symmetric coherent superposition 
of non-symmetric states. It is only in the thermodynamic limit when
the SSB may take place in a quantum system. In the case of the reggeon
field theory situation is even more complicated due to the fact that 
the evolution variable (the rapidity) would correspond to imaginary
time in the Schr\"{o}dinger picture. The consequences of the fact were 
investigated in detail in the framework of the reggeon field theory with
impact parameter dependence (RFT-b)~\cite{0dimb}. 
It turns out that in RFT-b the degenerate vacua communicate 
by quantum evolution irrespectively to the
extension of the system in the transverse space and the symmetry of 
quantum theory is maintained even in the thermodynamical limit. 
The communication was found to be realized by solitons in the impact 
parameter plane smoothly interpolating between the two asymmetric vacua.
In addition, the bifurcation of the classical solution at $Y=Y_c$ would 
indicate presence of a singularity of the $S$-matrix\footnote{We thank Lev Lipatov for this point.} in~$Y$. This singularity is not expected to 
be present in the complete quantum theory.

It should be stressed, however, that in this paper we do not address 
the issue of properties of the QCD pomeron field theory in the 
thermodynamical context. The goal is rather to get insight into 
scattering of two strong sources of colour field e.g.\ the nuclei.
In collisions of two nuclei the measurements give access not only to the
total cross sections but also to extended information about the kinematics
of the produced particles on the event-by-event basis. This is a classical
measurement which, necessarily, breaks the quantum coherence. Therefore it
is possible that such measurement selects one of the classical pomeron field 
trajectories which exhibit the symmetry breaking between the target 
and the projectile. In fact, the rapidity dependence of the saturation 
scale is different for the two asymmetric solutions of the Braun equations:
for one of the solutions the saturation scale increases from the target to
the projectile, while for the other it decreases. The average transverse 
momentum $\bar p_T$ of the emitted particles should be correlated with the 
saturation scale. Hence, a classical measurement of the event should 
select one of the asymmetric solutions and it could exhibit
some asymmetry in rapidity distribution of the produced particles.
The pattern may be somewhat obscured, however, when the dependence 
on the transverse position is taken into account. In the collision 
the regions separated in the impact parameter are only weakly correlated
and, in principle, it is possible that different domains in the transverse
space are dominated by different asymmetric solutions of Braun equations. 
This would make the possible effects of asymmetry more subtle and harder to 
disentangle.

One of the question which should be addressed is what observables could 
serve as experimental signatures of the asymmetry between the target 
and the projectile in heavy ion collisions. Certainly, the total cross section
carries no information about the details of the evolution, so one should 
focus on more detailed observables. As a first guess we would propose
investigation of the average 
transverse momentum~$\bar p_T = \sqrt{\langle p_T ^2 \rangle} $ 
of the particles produced in central collisions of heavy ions as a 
function of rapidity~$y$ in the c.m.s.\ frame on the event-by-event basis. 
With the symmetry between the target and the projectile being preserved 
the observable $\bar p_T(y)$ measured for individual events 
should be the same after changing the definition of rapidity $y \to -y$. 
If the symmetry is broken in the event, however, $\bar p_T(y)$ should 
exhibit a clear trend.

At this stage, we are not able to determine whether the symmetry breaking 
is a real physical phenomenon or an artifact of the effective theory of 
interacting pomerons. Needless to say, the framework of Braun equations 
relies on several assumptions that are far from being proven.
First of all, it is not clear if the pomerons are valid degrees of 
freedom in dense and strongly interacting gluonic systems. One may 
argue that at high density the pomerons overlap and melt down to 
gluons, whose dynamics may be significantly different from the
dynamics of the pomeron fields. Secondly, in the present analysis we
neglected quantum effects related to the pomeron loops. The impact
of the quantum effects on the phenomenon of symmetry breaking is 
unknown. Moreover, we neglected contribution of vertices with more than 
three pomerons. In addition, the NLL corrections to the BFKL pomeron kernels
and to the triple pomeron vertices are neglected in the present form
of Braun equations. This causes the BFKL intercept to be roughly two  
times too large. This means that the spontaneous breaking of the 
projectile-target symmetry should occur (if it occurs) at much higher 
rapidities than it may be deduced from the analysis employing the LL~BFKL 
kernel, perhaps for energies beyond reach of the LHC.  
On the other hand, the value of $\alpha_s=0.2$ underestimates 
significantly the expected value of $\alpha_s$ for the triple pomeron 
vertex (recall that the vertex is proportional to $\alpha_s^2$) and
with a more realistic value a smaller $Y_c$ would be predicted. 
Finally, in order to evaluate relevance of the effect the input
conditions should be carefully tuned to embody the available information
on the unintegrated gluon density in the nuclei, including the impact
parameter profiles. Keeping in mind all these reservation, we believe that 
further theoretical and experimental studies of the issue should
be carried out.

Leaving the issue of the symmetry breaking, we point out that 
the ``fan dominance'' phenomenon at $Y>Y_c$ found in the case of the 
symmetric boundary conditions should be even more pronounced when the 
projectile and the target are different. 
This might provide some basis for the use of the BK equation to 
describe the saturation effects in the DIS at low $Q^2$. 
Strictly speaking, the BK equation is valid for a small perturbative 
probe scattering off a large target, for instance a nucleus. This condition 
is, certainly, not fulfilled for almost real photon scattering off a proton,
nor it is for the diffractive DIS, dominated by scattering of large dipoles.
Still, the fits based on the BK amplitudes are very successful in both 
cases. The ``fan dominance'' in the symmetric Braun system could provide 
some support for those applications of the BK equation, although it is fair
to admit that the use of Braun equations to processes of this kind is not on 
the firm ground either.

The Braun equations are a minimal extension of the very 
fruitful concept of the BK equation, that embodies the symmetry between 
the pomeron fields $f$ and $\fda$ at the level of the action. 
Possibly, this extension may also find some interesting
phenomenological applications. Following Braun we state that description of 
heavy ion collisions is the obvious application of the equations. In that 
case the quantum loops should have only a subleading effect and the 
approximate treatment of the dependence transverse position is certainly 
sufficient.   A more challenging is an application of the formalism 
to the vital problem of understanding of $pp$ collisions at the LHC 
energies. In particular, we have in mind the description of underlying 
event, particle production (see e.g.~\cite{gsv,inew}), diffractive 
processes and 
determination of the gap survival factor in hard exclusive processes, 
like for instance the exclusive Higgs boson production. Expanding on this
example, the exclusive Higgs boson production is an important
process which is probable to be measured at the LHC~\cite{ehiggs}. 
It was shown, that the theoretical understanding of the cross section
for this process requires a good control of the hard rescattering 
corrections~\cite{higgsres}. 
The framework of the semi-classical field theory of the interacting 
pomerons may serve as a tool to perform the necessary resummations of
multi-pomeron diagrams and to obtain improved estimates of the gap
survival factor. All the listed applications are, however, non-trivial as
they require inclusion of NLL~BFKL corrections and, possibly, more accurate 
treatment of the dynamics of the system in the transverse position space.

In the last part of this section we will briefly mention some intriguing 
open questions. Thus, it would be interesting to investigate the stochastic 
QCD evolution of the color glass condensate using the realization 
of the semi-classical approximations in which only the tree topologies
of the pomeron are retained. In analogy to the Pomeron Field Theory, this
limit should be simpler than the accurate treatment also in the 
CGC formulation. Furthermore, a similar analysis of the Pomeron Field Theory 
should be possible after inclusion of the pomeron vertices at which
more than three pomeron fields. The form of those vertices may be predicted 
using the conjectured conformal symmetry of the pomeron field 
theory ~\cite{conformal}. 
While the conformal symmetry of the effective field theory of the interacting 
pomerons in the EGLLA was not explored in the present study, 
it constitutes, certainly, a key ingredient of the structure of the 
complete theory. Thus, it is mandatory to account for it in similar 
future studies.

Finally, let us refer to recent developments on the connection between
the superconformal gauge theories in four dimensions and the superstring 
theory on the $AdS_5\times S^5$ background~\cite{ads1}. Using the AdS/CFT 
duality it was found that the BFKL pomeron in gauge theories corresponds
to the graviton Regge trajectory in the AdS space~\cite{grav1}. 
Thus, it is desirable to find an interpretation of the effective pomeron field 
theory at the string side, perhaps in terms of gravity. 
Curiously enough, it was discovered recently that collisions of heavy 
ions possess a dual gravitational description~\cite{ion_gravity}. 
The dual of the scattering is given by a collision of two gravitational 
shock waves in which black holes can be formed. 
Example of such a black hole solution being produced, 
that moves in the fifth dimension of the Anti de Sitter space was 
found~\cite{ion_jp}. 
Thus, the relation between the fifth dimension in the AdS and the 
gluon virtuality inspires a question about the possible 
connection of the black hole to the BK traveling wave solution, 
whose existence we established in the classical pomeron field theory.
Therefore, it is important to verify whether phenomena analogous to
the symmetry breaking between the target and the projectile and the 
``fan dominance'' also happen in the string world.

\section{Conclusions}

Effective field theory of interacting QCD pomerons was investigated in the 
semi-classical limit, as a framework to describe high energy scattering of two
nuclei. The effective action was proposed in the form using the pomeron 
amplitudes, as the basic degrees of freedom, related to the 
unintegrated gluon densities in the linear regime.  Triple pomeron 
vertices in the momentum space accounted for the pomeron merging 
and splitting. Arbitrariness in the choice of the direction of evolution 
in rapidity required both the vertices to be identical and induced 
the self-duality of the action. 
This symmetry combined with symmetric initial conditions 
defined the scattering problem to be symmetric under the interchange of 
the target and the projectile. Classical pomeron field equations 
({\em Braun equations}) were re-derived and solved numerically. 
The solutions were found that were invariant under the projectile-target 
symmetry only for scattering rapidities~$Y$ smaller than a critical 
(non-universal) value $Y_c$. For $Y>Y_c$ the projectile-target symmetry 
turned out to be {\em spontaneously broken}. Above the critical rapidity, the 
solutions converged to the solutions of the Balitsky-Kovchegov equation,
the phenomenon which we called the {\em BK fan dominance}. A very similar 
pattern of symmetry breaking and the fan dominance occurs also in the 
Reggeon Field Theory in zero transverse dimensions, which suggests that this
is a generic feature of the interacting pomeron system. 
We discussed possible consequences of those observations for the 
phenomenology of heavy ion collisions and the physics of $pp$ scattering at
the LHC. Finally, we suggested that the results of this paper may have
counterparts in the dual description of heavy ion collisions in terms of
scattering of two gravitational shock waves in the Anti de~Sitter space in
five dimensions.

\section*{Acknowledgments}
We are especially grateful to Jochen Bartels for his continued 
interest in this work and numerous enlightening discussions. 
We thank Mikhail Braun, Krzysztof Golec-Biernat, Eugene Levin, Lev Lipatov 
and Alfred Mueller for discussions and useful comments.  
S.B.\ thanks the Minerva foundation for its support and
L.M.\ gratefully acknowledges the support of the grant 
of the Polish State Committee for Scientific Research 
No.\ 1~P03B~028~28.


\vspace{3mm}


\begin{thebibliography}{99}

\bibitem{ads1}
  J.~M.~Maldacena,
  %``The large N limit of superconformal field theories and supergravity,''
  Adv.\ Theor.\ Math.\ Phys.\  {\bf 2} (1998) 231
  [Int.\ J.\ Theor.\ Phys.\  {\bf 38} (1999) 1113];
%  [arXiv:hep-th/9711200].
  %%CITATION = HEP-TH 9711200;%%
%\bibitem{ads2}
  E.~Witten,
  %``Anti-de Sitter space and holography,''
  Adv.\ Theor.\ Math.\ Phys.\  {\bf 2} (1998) 253;
%  [arXiv:hep-th/9802150].
  %%CITATION = HEP-TH 9802150;%%
%
%\bibitem{ads3}
  O.~Aharony, S.~S.~Gubser, J.~M.~Maldacena, H.~Ooguri and Y.~Oz,
  %``Large N field theories, string theory and gravity,''
  Phys.\ Rept.\  {\bf 323} (2000) 183.
%  [arXiv:hep-th/9905111].
  %%CITATION = HEP-TH 9905111;%%


%%%%%%%%%%%%%%%%%%%%%

\bibitem{bfkl}
  L.~N.~Lipatov,
  %``Reggeization Of The Vector Meson And The Vacuum Singularity In Nonabelian
  %Gauge Theories,''
  Sov.\ J.\ Nucl.\ Phys.\  {\bf 23} (1976) 338
  [Yad.\ Fiz.\  {\bf 23} (1976) 642];
  %%CITATION = SJNCA,23,338;%%
%
%\bibitem{bfkl2}
  E.~A.~Kuraev, L.~N.~Lipatov and V.~S.~Fadin,
  %``The Pomeranchuk Singularity In Nonabelian Gauge Theories,''
  Sov.\ Phys.\ JETP {\bf 45} (1977) 199
  [Zh.\ Eksp.\ Teor.\ Fiz.\  {\bf 72} (1977) 377];
  %%CITATION = SPHJA,45,199;%%
%
%\bibitem{bfkl3}
  I.~I.~Balitsky and L.~N.~Lipatov,
  %``The Pomeranchuk Singularity In Quantum Chromodynamics,''
  Sov.\ J.\ Nucl.\ Phys.\  {\bf 28} (1978) 822
  [Yad.\ Fiz.\  {\bf 28} (1978) 1597].
  %%CITATION = SJNCA,28,822;%%

\bibitem{bfklsum}
L.~N.~Lipatov, {Phys.\ Rept.\ } {\bf  286} (1997) 131.

%\cite{Fadin:1998py}
\bibitem{nlbfkl}
  V.~S.~Fadin and L.~N.~Lipatov,
  %``BFKL Pomeron in the next-to-leading approximation,''
  Phys.\ Lett.\ B {\bf 429} (1998) 127;
%  %[arXiv:hep-ph/9802290].
  %%CITATION = HEP-PH 9802290;%%
%
%\cite{Ciafaloni:1998gs}
%\bibitem{nlbfkl2}
M.~Ciafaloni and G.~Camici,
%``Energy scale(s) and next-to-leading BFKL equation,''
  Phys.\ Lett.\ B {\bf 430} (1998) 349;
%  %[arXiv:hep-ph/9803389].
  %%CITATION = HEP-PH 9803389;%%
%
%\cite{Fadin:2004zq}
%\bibitem{nlbfkl3}
  V.~S.~Fadin and R.~Fiore,
  %``Non-forward BFKL pomeron at next-to-leading order,''
  Phys.\ Lett.\ B {\bf 610} (2005) 61
  [Erratum-ibid.\ B {\bf 621} (2005) 61];
  %[arXiv:hep-ph/0412386].
  %%CITATION = HEP-PH 0412386;%%
%
%\cite{Fadin:2005zj}
%\bibitem{nlbfkl4}
  V.~S.~Fadin and R.~Fiore,
  %``Non-forward NLO BFKL kernel,''
  Phys.\ Rev.\ D {\bf 72} (2005) 014018.
  %[arXiv:hep-ph/0502045].
  %%CITATION = HEP-PH 0502045;%%

%%%%%%%%%%%%%%%%%%%%%%%%%%


%\cite{Balitsky:1995ub}
\bibitem{balitsky}
  I.~Balitsky,
  %``Operator expansion for high-energy scattering,''
  Nucl.\ Phys.\ B {\bf 463} (1996) 99.
%  [arXiv:hep-ph/9509348].
  %%CITATION = HEP-PH 9509348;%%

%

%\cite{Jalilian-Marian:1997dw}
\bibitem{jimwalk}
  J.~Jalilian-Marian, A.~Kovner and H.~Weigert,
  %``The Wilson renormalization group for low x physics: Gluon evolution at
  %finite parton density,''
  Phys.\ Rev.\ D {\bf 59} (1999) 014015;
  %[arXiv:hep-ph/9709432].
  %%CITATION = HEP-PH 9709432;%%
%
%\cite{Jalilian-Marian:1997gr}
%\bibitem{Jalilian-Marian:1997gr}
J.~Jalilian-Marian, A.~Kovner, A.~Leonidov and H.~Weigert,
%``The Wilson renormalization group for low x physics: Towards the high
%density regime,''
Phys.\ Rev.\ D {\bf 59} (1999) 014014;
  %[arXiv:hep-ph/9706377].
  %%CITATION = HEP-PH 9706377;%%
%
%\cite{Iancu:2000hn}
%\bibitem{Iancu:2000hn}
  E.~Iancu, A.~Leonidov and L.~D.~McLerran,
  %``Nonlinear gluon evolution in the color glass condensate. I,''
  Nucl.\ Phys.\ A {\bf 692} (2001) 583;
  %[arXiv:hep-ph/0011241].
  %%CITATION = HEP-PH 0011241;%%
%
%\cite{Iancu:2001ad}
%\bibitem{Iancu:2001ad}
  E.~Iancu, A.~Leonidov and L.~D.~McLerran,
  %``The renormalization group equation for the color glass condensate,''
  Phys.\ Lett.\ B {\bf 510} (2001) 133;
  %[arXiv:hep-ph/0102009].
  %%CITATION = HEP-PH 0102009;%%
%  
%\cite{Iancu:2001md}
%\bibitem{Iancu:2001md}
  E.~Iancu and L.~D.~McLerran,
  %``Saturation and universality in QCD at small x,''
  Phys.\ Lett.\ B {\bf 510} (2001) 145;
  %[arXiv:hep-ph/0103032].
  %%CITATION = HEP-PH 0103032;%%
%
%\cite{Ferreiro:2001qy}
%\bibitem{Ferreiro:2001qy}
  E.~Ferreiro, E.~Iancu, A.~Leonidov and L.~McLerran,
  %``Nonlinear gluon evolution in the color glass condensate. II,''
  Nucl.\ Phys.\ A {\bf 703} (2002) 489.
  %[arXiv:hep-ph/0109115].
  %%CITATION = HEP-PH 0109115;%%

%%%%%%%%%%%%%%%%%%%%%%%%%%



%\cite{Mueller:1993rr}
\bibitem{dipmod}
  A.~H.~Mueller,
  %``Soft gluons in the infinite momentum wave function and the BFKL pomeron,''
  Nucl.\ Phys.\ B {\bf 415} (1994) 373.
  %%CITATION = NUPHA,B415,373;%%

%\cite{Iancu:2004iy}
\bibitem{stochastic}
%\bibitem{Iancu:2004es}
  E.~Iancu, A.~H.~Mueller and S.~Munier,
  %``Universal behavior of QCD amplitudes at high energy from general tools  of
  %statistical physics,''
  Phys.\ Lett.\ B {\bf 606} (2005) 342;
%[arXiv:hep-ph/0410018].
%%CITATION = HEP-PH 0410018;%%
%
  E.~Iancu and D.~N.~Triantafyllopoulos,
  %``A Langevin equation for high energy evolution with pomeron loops,''
  Nucl.\ Phys.\ A {\bf 756} (2005) 419;
  %[arXiv:hep-ph/0411405].
  %%CITATION = HEP-PH 0411405;%%
%
%\cite{Mueller:2005ut}
%\bibitem{Mueller:2005ut}
  A.~H.~Mueller, A.~I.~Shoshi and S.~M.~H.~Wong,
  %``Extension of the JIMWLK equation in the low gluon density region,''
  Nucl.\ Phys.\ B {\bf 715} (2005) 440;
  %[arXiv:hep-ph/0501088].
  %%CITATION = HEP-PH 0501088;%%
%
%
%\cite{Munier:2005re}
%\bibitem{Munier:2005re}
  S.~Munier,
  %``High energy scattering in QCD as a statistical process,''
  Nucl.\ Phys.\ A {\bf 755} (2005) 622;
%  [arXiv:hep-ph/0501149].
  %%CITATION = HEP-PH 0501149;%%
%
E.~Levin and M.~Lublinsky,
  %``Towards a symmetric approach to high energy evolution: Generating
  %functional with Pomeron loops,''
  Nucl.\ Phys.\ A {\bf 763} (2005) 172;
% [arXiv:hep-ph/0501173].
  %%CITATION = HEP-PH 0501173;%%
%
%\cite{Iancu:2005nj}
%\bibitem{Iancu:2005nj}
E.~Iancu and D.~N.~Triantafyllopoulos,
%``Non-linear QCD evolution with improved triple-pomeron vertices,''
Phys.\ Lett.\ B {\bf 610} (2005) 253;
%[arXiv:hep-ph/0501193].
%%CITATION = HEP-PH 0501193;%%
%
%
%\cite{Enberg:2005cb}
%\bibitem{Enberg:2005cb}
  R.~Enberg, K.~Golec-Biernat and S.~Munier,
  %``The high energy asymptotics of scattering processes in QCD,''
  Phys.\ Rev.\ D {\bf 72} (2005) 074021;
%  [arXiv:hep-ph/0505101].
%%CITATION = HEP-PH 0505101;%%
%
%\bibitem{Triantafyllopoulos:2005cn}
  D.~N.~Triantafyllopoulos,
  %``Pomeron loops in high energy QCD,''
  Acta Phys.\ Polon.\ B {\bf 36} (2005) 3593.
  %[arXiv:hep-ph/0511226].
  %%CITATION = HEP-PH 0511226;%%



\bibitem{vert1}
J.~Bartels,
%``Unitarity corrections to the Lipatov pomeron and the four gluon operator in
%deep inelastic scattering in QCD,''
Z.\ Phys.\ C {\bf 60} (1993) 471.
%%CITATION = ZEPYA,C60,471;%%

\bibitem{vert2}
J.~Bartels and M.~W\"{u}sthoff,
%``The Triple Regge limit of diffractive dissociation in deep inelastic
%scattering,''
Z.\ Phys.\ C {\bf 66} (1995) 157.
%%CITATION = ZEPYA,C66,157;%%


\bibitem{eglla1}
  J.~Bartels and C.~Ewerz,
  %``Unitarity corrections in high-energy QCD,''
  JHEP {\bf 9909} (1999) 026.
  %[arXiv:hep-ph/9908454].
  %%CITATION = HEP-PH 9908454;%%


%\cite{Ewerz:2001uq}
\bibitem{eglla2}
  C.~Ewerz,
  %``Conformal invariance of unitarity corrections,''
  Phys.\ Lett.\ B {\bf 512} (2001) 239.
  %[arXiv:hep-ph/0105181].
  %%CITATION = HEP-PH 0105181;%%


%\cite{Ewerz:2003an}
\bibitem{eglla3}
  C.~Ewerz and V.~Schatz,
  %``How pomerons meet in coloured glass,''
  Nucl.\ Phys.\ A {\bf 736} (2004) 371.
  %[arXiv:hep-ph/0308056].
  %%CITATION = HEP-PH 0308056;%%


%\cite{Bittig:2005ni}
\bibitem{eglla4}
  T.~Bittig and C.~Ewerz,
  %``Diffraction, the color glass condensate and string theory,''
  Nucl.\ Phys.\ A {\bf 755} (2005) 616.
%[arXiv:hep-ph/0501192].
 %%CITATION = HEP-PH 0501192;%%


%\cite{Kovchegov:1999yj}
\bibitem{kov1}
  Y.~V.~Kovchegov,
  %``Small-x F2 structure function of a nucleus including multiple Pomeron
  %exchanges,''
  Phys.\ Rev.\ D {\bf 60} (1999) 034008.
%  [arXiv:hep-ph/9901281].
  %%CITATION = HEP-PH 9901281;%%

%\cite{Kovchegov:1999ua}
\bibitem{kov2}
  Y.~V.~Kovchegov,
  %``Unitarization of the BFKL Pomeron on a nucleus,''
  Phys.\ Rev.\ D {\bf 61} (2000) 074018.
%  [arXiv:hep-ph/9905214].
  %%CITATION = HEP-PH 9905214;%%



%cite{GBW}
\bibitem{gbw1}
  K.~Golec-Biernat and M.~W\"{u}sthoff,
  %``Saturation effects in deep inelastic scattering at low Q**2 and its
  %implications on diffraction,''
  Phys.\ Rev.\ D {\bf 59} (1999) 014017.
 % [arXiv:hep-ph/9807513].
  %%CITATION = HEP-PH 9807513;%%

%cite{GBW2}
\bibitem{gbw2}
K.~Golec-Biernat and M.~W\"{u}sthoff,
  %``Saturation in diffractive deep inelastic scattering,''
  Phys.\ Rev.\ D {\bf 60} (1999) 114023.
%  [arXiv:hep-ph/9903358].
  %%CITATION = HEP-PH 9903358;%%


%%%%%%%%%%%%%%%%%%%%%%%%%%%%
%%%%braun


%BK sols


%\cite{Levin:1999mw}
\bibitem{bksols}
  E.~Levin and K.~Tuchin,
  %``Solution to the evolution equation for high parton density QCD,''
  Nucl.\ Phys.\ B {\bf 573} (2000) 833;
  %[arXiv:hep-ph/9908317].
  %%CITATION = HEP-PH 9908317;%%
%
%\cite{Braun:2000wr}
%\bibitem{Braun:2000wr}
M.~Braun,
%``Structure function of the nucleus in the perturbative QCD with N(c)  $\to$
%infinity (BFKL pomeron fan diagrams),''
Eur.\ Phys.\ J.\ C {\bf 16} (2000) 337;
%[arXiv:hep-ph/0001268].
%%CITATION = HEP-PH 0001268;%%
%
%\cite{Weigert:2000gi}
%\bibitem{Weigert:2000gi}
  H.~Weigert,
  %``Unitarity at small Bjorken x,''
  Nucl.\ Phys.\ A {\bf 703} (2002) 823;
%  [arXiv:hep-ph/0004044];
  %%CITATION = HEP-PH 0004044;%%
%\bibitem{Lublinsky:2001yi}
  M.~Lublinsky, E.~Gotsman, E.~Levin and U.~Maor,
%``Non-linear evolution and parton distributions at LHC and THERA energies,''
  Nucl.\ Phys.\ A {\bf 696} (2001) 851;
%[arXiv:hep-ph/0102321].
%%CITATION = HEP-PH 0102321;%%
%
%\bibitem{AB}
  N.~Armesto and M.~A.~Braun,
  %``Parton densities and dipole cross-sections at small x in large nuclei,''
  Eur.\ Phys.\ J.\ C {\bf 20} (2001) 517;
%  [arXiv:hep-ph/0104038].
  %%CITATION = HEP-PH 0104038;%%
%
%\bibitem{GBMS}
  K.~Golec-Biernat, L.~Motyka and A.~M.~Sta\'{s}to,
  %``Diffusion into infrared and unitarization of the BFKL Pomeron,''
  Phys.\ Rev.\ D {\bf 65} (2002) 074037;
%  [arXiv:hep-ph/0110325].
  %%CITATION = HEP-PH 0110325;%%
%\cite{Chachamis:2004ab}
%\bibitem{Chachamis:2004ab}
G.~Chachamis, M.~Lublinsky and A.~Sabio Vera,
%``Higher order effects in non linear evolution from a veto in rapidities,''
Nucl.\ Phys.\ A {\bf 748} (2005) 649.
%[arXiv:hep-ph/0408333].
%%CITATION = HEP-PH 0408333;%%
%


\bibitem{bdepkov}
  K.~Golec-Biernat and A.~M.~Sta\'{s}to,
  %``On solutions of the Balitsky-Kovchegov equation with impact parameter,''
  Nucl.\ Phys.\ B {\bf 668} (2003) 345.
  %[arXiv:hep-ph/0306279].
  %%CITATION = HEP-PH 0306279;%%

\bibitem{jimsol}
  K.~Rummukainen and H.~Weigert,
  %``Universal features of JIMWLK and BK evolution at small x,''
  Nucl.\ Phys.\ A {\bf 739} (2004) 183.
  %[arXiv:hep-ph/0309306].
  %%CITATION = HEP-PH 0309306;%%

%%%%%%%%%%%%%%%


%\cite{Levin:2000mv}
\bibitem{bksemi1}
  E.~Levin and K.~Tuchin,
  %``New scaling at high energy DIS,''
  Nucl.\ Phys.\ A {\bf 691} (2001) 779;
  %[arXiv:hep-ph/0012167].
  %%CITATION = HEP-PH 0012167;%%
%
%\cite{Kwiecinski:2002ep}
%\bibitem{Kwiecinski:2002ep}
  J.~Kwieci\'{n}ski and A.~M.~Sta\'{s}to,
  %``Geometric scaling and QCD evolution,''
  Phys.\ Rev.\ D {\bf 66} (2002) 014013;
  %[arXiv:hep-ph/0203030].
  %%CITATION = HEP-PH 0203030;%%
%
%
%\cite{Iancu:2002tr}
  E.~Iancu, K.~Itakura and L.~McLerran,
  %``Geometric scaling above the saturation scale,''
  Nucl.\ Phys.\ A {\bf 708} (2002) 327.
  %[arXiv:hep-ph/0203137].
  %%CITATION = HEP-PH 0203137;%%


%\cite{Mueller:2002zm}
\bibitem{bksemi2}
  A.~H.~Mueller and D.~N.~Triantafyllopoulos,
  %``The energy dependence of the saturation momentum,''
  Nucl.\ Phys.\ B {\bf 640} (2002) 331;
  %[arXiv:hep-ph/0205167].
  %%CITATION = HEP-PH 0205167;%%
%
%\cite{Triantafyllopoulos:2002nz}
%\bibitem{Triantafyllopoulos:2002nz}
  D.~N.~Triantafyllopoulos,
  %``The energy dependence of the saturation momentum from RG improved BFKL
  %evolution,''
  Nucl.\ Phys.\ B {\bf 648} (2003) 293;
  %[arXiv:hep-ph/0209121].
  %%CITATION = HEP-PH 0209121;%%
%\cite{Motyka:2005ep}
%\bibitem{lmodd}
  L.~Motyka, Phys.\ Lett.\ B, in print;
  %``Nonlinear evolution of pomeron and odderon in momentum space,''
  arXiv:hep-ph/0509270.
%%CITATION = HEP-PH 0509270;%%
%%%%%%%%%%%%%%%%%%%%%





%\cite{Gotsman:2002yy}
\bibitem{bk-pheno}
E.~Gotsman, E.~Levin, M.~Lublinsky and U.~Maor,
%``Towards a new global QCD analysis: Low x DIS data from non-linear
%evolution,''
Eur.\ Phys.\ J.\ C {\bf 27} (2003) 411;
%[arXiv:hep-ph/0209074].
%%CITATION = HEP-PH 0209074;%%
%
%


\bibitem{kks}
K.~Kutak and J.~Kwieci\'{n}ski,
%``Screening effects in the ultrahigh energy neutrino interactions,''
Eur.\ Phys.\ J.\ C {\bf 29} (2003) 521;
%[arXiv:hep-ph/0303209].
%%CITATION = HEP-PH 0303209;%%
%\cite{Kutak:2004ym}
%\bibitem{Kutak:2004ym}
K.~Kutak and A.~M.~Sta\'{s}to,
%``Unintegrated gluon distribution from modified BK equation,''
Eur.\ Phys.\ J.\ C {\bf 41} (2005) 343.
%[arXiv:hep-ph/0408117].
%%CITATION = HEP-PH 0408117;%%


\bibitem{iim}
E.~Iancu, K.~Itakura and S.~Munier,
%``Saturation and BFKL dynamics in the HERA data at small x,''
Phys.\ Lett.\ B {\bf 590} (2004) 199.
%[arXiv:hep-ph/0310338].
%

%%%%%%%%%%






%scaling

\bibitem{scaling}
  A.~M.~Sta\'{s}to, K.~Golec-Biernat and J.~Kwieci\'{n}ski,
  %``Geometric scaling for the total gamma* p cross-section in the low x
  %region,''
  Phys.\ Rev.\ Lett.\  {\bf 86} (2001) 596.
  %[arXiv:hep-ph/0007192].
  %%CITATION = HEP-PH 0007192;%%

\bibitem{barlev}
J.~Bartels and E.~Levin,
%``Solutions to the Gribov-Levin-Ryskin equation in the nonperturbative
%region,''
Nucl.\ Phys.\ B {\bf 387} (1992) 617.
%%CITATION = NUPHA,B387,617;%%


\bibitem{traveling}
  S.~Munier and R.~Peschanski,
  %``Geometric scaling as traveling waves,''
  Phys.\ Rev.\ Lett.\  {\bf 91} (2003) 232001;
%  [arXiv:hep-ph/0309177].
  %%CITATION = HEP-PH 0309177;%%
%
%\bibitem{MuPe2}
  S.~Munier and R.~Peschanski,
  %``Traveling wave fronts and the transition to saturation,''
  Phys.\ Rev.\ D {\bf 69}, 034008 (2004);
% [arXiv:hep-ph/0310357].
  %%CITATION = HEP-PH 0310357;%%
%
%\bibitem{MuPe3}
  S.~Munier and R.~Peschanski,
  %``Universality and tree structure of high energy QCD,''
  Phys.\ Rev.\ D {\bf 70}, 077503 (2004).
%  [arXiv:hep-ph/0401215].
  %%CITATION = HEP-PH 0401215;%%


%%%%%%%%%%%%%%%%%%%%%


%\cite{Kovchegov:2003dm}
%\bibitem{whimiks}
%  Y.~V.~Kovchegov, L.~Szymanowski and S.~Wallon,
%  %``Perturbative odderon in the dipole model,''
%  Phys.\ Lett.\ B {\bf 586} (2004) 267;
%  %[arXiv:hep-ph/0309281].
%  %%CITATION = HEP-PH 0309281;%%
%
%%\cite{Hatta:2005as}
%%\bibitem{Hatta:2005as}
%  Y.~Hatta, E.~Iancu, K.~Itakura and L.~McLerran,
%  %``Odderon in the color glass condensate,''
%  Nucl.\ Phys.\ A {\bf 760} (2005) 172.
%  %[arXiv:hep-ph/0501171].
%  %%CITATION = HEP-PH 0501171;%%


%\cite{Motyka:2005ep}
%\bibitem{lmodd}
%  L.~Motyka, Phys.\ Lett.\ B, in print;
%  %``Nonlinear evolution of pomeron and odderon in momentum space,''
%  arXiv:hep-ph/0509270.
%  %%CITATION = HEP-PH 0509270;%%

%%%%%%%%%%%%%%%


%\cite{Bartels:2002cj}
\bibitem{gbwev}
  J.~Bartels, K.~Golec-Biernat and H.~Kowalski,
  %``A modification of the saturation model: DGLAP evolution,''
  Phys.\ Rev.\ D {\bf 66} (2002) 014001.
  %[arXiv:hep-ph/0203258].
  %%CITATION = HEP-PH 0203258;%%
%

%\cite{Timneanu:2001bk}
\bibitem{tkm}
  N.~T\^{i}mneanu, J.~Kwieci\'{n}ski and L.~Motyka,
  %``Saturation model for two-photon interactions at high energies,''
  Eur.\ Phys.\ J.\ C {\bf 23} (2002) 513.
  %[arXiv:hep-ph/0110409].
  %%CITATION = HEP-PH 0110409;%%
%
%\cite{Kowalski:2003hm}
\bibitem{kt}
  H.~Kowalski and D.~Teaney,
  %``An impact parameter dipole saturation model,''
  Phys.\ Rev.\ D {\bf 68} (2003) 114005.
  %[arXiv:hep-ph/0304189].
  %%CITATION = HEP-PH 0304189;%%
%
\bibitem{kmw}
H.~Kowalski, L.~Motyka and G.~Watt, in preparation; K.~Kowalski, talk given
at XIV International Workshop on Deep Inelastic Scattering, April 2006, 
Tsukuba, Japan.
%



\bibitem{braun1}
%\cite{Braun:2000bi}
%\bibitem{Braun:2000bi}
  M.~A.~Braun,
  %``Nucleus nucleus scattering in perturbative QCD with N(c) $\to$ infinity,''
  Phys.\ Lett.\ B {\bf 483} (2000) 115.
%   [arXiv:hep-ph/0003004].
  %%CITATION = HEP-PH 0003004;%%

%\cite{Braun:2003dz}
\bibitem{braun2}
  M.~A.~Braun,
  %``Nucleus nucleus interaction in the perturbative QCD,''
  Eur.\ Phys.\ J.\ C {\bf 33} (2004) 113.
  %[arXiv:hep-ph/0309293].
  %%CITATION = HEP-PH 0309293;%%

%\cite{Braun:2005hx}
\bibitem{braun3}
  M.~A.~Braun,
  %``Conformal invariant pomeron interaction in the perurbative QCD with large
  %N(c),''
  Phys.\ Lett.\ B {\bf 632} (2006) 297.
  %[arXiv:hep-ph/0512057].
  %%CITATION = HEP-PH 0512057;%%


%%%%%%%%%%%%%%%%%%%



%%%two boundaries
%\cite{Mueller:2004se}
\bibitem{absorptive}
  A.~H.~Mueller and A.~I.~Shoshi,
  %``Small-x physics beyond the Kovchegov equation,''
  Nucl.\ Phys.\ B {\bf 692} (2004) 175.
  %[arXiv:hep-ph/0402193].
  %%CITATION = HEP-PH 0402193;%%


%balitsky

%\cite{Balitsky:2004rr}
\bibitem{balshock}
  I.~Balitsky,
  %``Scattering of shock waves in QCD,''
  Phys.\ Rev.\ D {\bf 70} (2004) 114030;
  %[arXiv:hep-ph/0409314].
  %%CITATION = HEP-PH 0409314;%%
%
%\cite{Balitsky:2005we}
%\bibitem{balshock2}
  I.~Balitsky,
  %``High-enegy effective action from scattering of QCD shock waves,''
  Phys.\ Rev.\ D {\bf 72} (2005) 074027.
  %[arXiv:hep-ph/0507237].
  %%CITATION = HEP-PH 0507237;%%






%%%%%%%%%%%%%%%%%%%%%%%%%%%%%%%
%0-dim


\bibitem{0dimc}
  D.~Amati, L.~Caneschi and R.~Jengo,
  %``Summing Pomeron Trees,''
  Nucl.\ Phys.\ B {\bf 101} (1975) 397.
  %%CITATION = NUPHA,B101,397;%%
%
%\cite{Jengo:1976nt}

\bibitem{0dimq}
  R.~Jengo,
  %``Zero Slope Limit Of The Pomeron Field Theory,''
  Nucl.\ Phys.\ B {\bf 108} (1976) 447;
  %%CITATION = NUPHA,B108,447;%%
%
%\cite{Amati:1976ck}
%\bibitem{Amati:1976ck}
  D.~Amati, M.~Le Bellac, G.~Marchesini and M.~Ciafaloni,
  %``Reggeon Field Theory For Alpha (O) > 1,''
  Nucl.\ Phys.\ B {\bf 112} (1976) 107;
  %%CITATION = NUPHA,B112,107;%%
%
%\bibitem{Ciafaloni:1977xv}
  M.~Ciafaloni, M.~Le Bellac and G.~C.~Rossi,
  %``Reggeon Quantum Mechanics: A Critical Discussion,''
  Nucl.\ Phys.\ B {\bf 130} (1977) 388.
  %%CITATION = NUPHA,B130,388;%%

\bibitem{0dimi}
  M.~Ciafaloni,
  %``Instanton Contributions In Reggeon Quantum Mechanics,''
  Nucl.\ Phys.\ B {\bf 146} (1978) 427.
  %%CITATION = NUPHA,B146,427;%%


%self-duality

%\cite{Kovner:2005en}
\bibitem{selfdual}
A.~Kovner and M.~Lublinsky,
%``From target to projectile and back again: Selfduality of high energy
%evolution,''
Phys.\ Rev.\ Lett.\  {\bf 94} (2005) 181603;
%[arXiv:hep-ph/0502119].
%%CITATION = HEP-PH 0502119;%%
%
%\cite{Blaizot:2005vf}
%\bibitem{self2}
  J.~P.~Blaizot, E.~Iancu, K.~Itakura and D.~N.~Triantafyllopoulos,
  %``Duality and Pomeron effective theory for QCD at high energy and large
  %N(c),''
  Phys.\ Lett.\ B {\bf 615} (2005) 221;
  %[arXiv:hep-ph/0502221].
  %%CITATION = HEP-PH 0502221;%%
%
%\bibitem{self3}
Y.~Hatta, E.~Iancu, L.~McLerran, A.~Sta\'{s}to and D.~N.~Triantafyllopoulos,
%``Effective Hamiltonian for QCD evolution at high energy,''
Nucl.\ Phys.\ A {\bf 764} (2006) 423;
%[arXiv:hep-ph/0504182].
%%CITATION = HEP-PH 0504182;%%
%
%
%\cite{Marquet:2005hu}
%\bibitem{self4}
  C.~Marquet, A.~H.~Mueller, A.~I.~Shoshi and S.~M.~H.~Wong,
  %``On the projectile-target duality of the color glass condensate in the
  %dipole picture,''
  Nucl.\ Phys.\ A {\bf 762} (2005) 252.
  %[arXiv:hep-ph/0505229].
  %%CITATION = HEP-PH 0505229;%%



% froissart 

%\cite{Kovner:2001bh}
\bibitem{kovn1}
  A.~Kovner and U.~A.~Wiedemann,
  %``Nonlinear QCD evolution: Saturation without unitarization,''
  Phys.\ Rev.\ D {\bf 66} (2002) 051502;
%  [arXiv:hep-ph/0112140].
  %%CITATION = HEP-PH 0112140;%%
%
%\bibitem{kovn2}
A.~Kovner and U.~A.~Wiedemann,
%``No Froissart bound from gluon saturation,''
Phys.\ Lett.\ B {\bf 551} (2003) 311.
%  [arXiv:hep-ph/0207335].
  %%CITATION = HEP-PH 0207335;%%



%\cite{0dimnew}
\bibitem{0dimnew}
P.~Rembiesa and A.~M.~Sta\'{s}to,
%``Algebraic models for the hierarchy structure of evolution equations at small
%x,''
Nucl.\ Phys.\ B {\bf 725} (2005) 251;
%[arXiv:hep-ph/0503223].
%%CITATION = HEP-PH 0503223;%%
%
%\cite{Shoshi:2005pf}
%\bibitem{Shoshi:2005pf}
A.~I.~Shoshi and B.~W.~Xiao,
%``Pomeron loops in zero transverse dimensions,''
  Phys.\ Rev.\ D {\bf 73} (2006) 094014;
%  [arXiv:hep-ph/0512206].
  %%CITATION = HEP-PH 0512206;%%
%\cite{Shoshi:2006eb}
%\bibitem{Shoshi:2006eb}
  A.~I.~Shoshi and B.~W.~Xiao,
  %``Diffractive dissociation including pomeron loops in zero transverse
  %dimensions,''
  arXiv:hep-ph/0605282.
  %%CITATION = HEP-PH 0605282;%%


%
%\cite{Kozlov:2006zj}
%
\bibitem{0dimkl}
  M.~Kozlov and E.~Levin,
  %``Solution for the BFKL pomeron calculus in zero transverse dimensions,''
  arXiv:hep-ph/0604039.
  %%CITATION = HEP-PH 0604039;%%




\bibitem{0dimour}
S.~Bondarenko and L.~Motyka, in preparation.



\bibitem{0dimb}
  D.~Amati, G.~Marchesini, M.~Ciafaloni and G.~Parisi,
  %``Expanding Disk As A Dynamical Vacuum Instability In Reggeon Field Theory,''
  Nucl.\ Phys.\ B {\bf 114} (1976) 483;
  %%CITATION = NUPHA,B114,483;%%
  V.~Alessandrini, D.~Amati and M.~Ciafaloni,
  %``Classical Kinks And Their Quantization In Supercritical Reggeon Field
  %Theory,''
  Nucl.\ Phys.\ B {\bf 130} (1977) 429.
  %%CITATION = NUPHA,B130,429;%%





\bibitem{gsv}
  F.~Gelis, A.~M.~Sta\'{s}to and R.~Venugopalan,
  %``Limiting fragmentation in hadron-hadron collisions at high energies,''
  arXiv:hep-ph/0605087.
  %%CITATION = HEP-PH 0605087;%%

\bibitem{inew}
  E.~Iancu, C.~Marquet and G.~Soyez,
  %``Forward gluon production in hadron-hadron scattering with Pomeron loops,''
  arXiv:hep-ph/0605174.
  %%CITATION = HEP-PH 0605174;%%


%\cite{Khoze:2000cy}
\bibitem{ehiggs}
  V.~A.~Khoze, A.~D.~Martin and M.~G.~Ryskin,
  %``Can the Higgs be seen in rapidity gap events at the Tevatron or the
  %LHC?,''
  Eur.\ Phys.\ J.\ C {\bf 14} (2000) 525;
%  [arXiv:hep-ph/0002072].
  %%CITATION = HEP-PH 0002072;%%
%\cite{DeRoeck:2002hk}
%\bibitem{DeRoeck:2002hk}
  A.~De Roeck, V.~A.~Khoze, A.~D.~Martin, R.~Orava and M.~G.~Ryskin,
  %``Ways to detect a light Higgs boson at the LHC,''
  Eur.\ Phys.\ J.\ C {\bf 25} (2002) 391;
%  [arXiv:hep-ph/0207042].
  %%CITATION = HEP-PH 0207042;%%
%\cite{Kaidalov:2003fw}
%\bibitem{Kaidalov:2003fw}
  A.~B.~Kaidalov, V.~A.~Khoze, A.~D.~Martin and M.~G.~Ryskin,
  %``Central exclusive diffractive production as a spin parity analyser:  From
  %hadrons to Higgs,''
  Eur.\ Phys.\ J.\ C {\bf 31} (2003) 387.
%  [arXiv:hep-ph/0307064].
  %%CITATION = HEP-PH 0307064;%%


%\cite{Bartels:2006ea}
\bibitem{higgsres}
  J.~Bartels, S.~Bondarenko, K.~Kutak and L.~Motyka, 
  Phys.\ Rev.\ D {\bf 73} (2006) 093004;
  %``Exclusive Higgs boson production at the LHC: Hard rescattering
  %corrections,''
  arXiv:hep-ph/0601128.
  %%CITATION = HEP-PH 0601128;%%




\bibitem{conformal}
%\bibitem{3p-conf}
J.~Bartels, L.~N.~Lipatov and M.~W\"{u}sthoff,
  %``Conformal Invariance of the Transition Vertex $2 \to 4$ gluons,''
Nucl.\ Phys.\ B {\bf 464} (1996) 298;
%  %[arXiv:hep-ph/9509303].
%  %%CITATION = HEP-PH 9509303;%%
%%%%%%%%%
% equivalnece of the vertices
G.~P.~Korchemsky,
%``Conformal bootstrap for the BFKL pomeron,''
Nucl.\ Phys.\ B {\bf 550} (1999) 397;
%%[arXiv:hep-ph/9711277].
%%%CITATION = HEP-PH 9711277;%%
%\cite{3p_bv}
%\bibitem{3p_bv}
M.~A.~Braun and G.~P.~Vacca,
%``Triple pomeron vertex in the limit N(c) $\to$ infinity,''
Eur.\ Phys.\ J.\ C {\bf 6} (1999) 147;
%[arXiv:hep-ph/9711486].
%%%CITATION = HEP-PH 9711486;%%
%
%%\cite{Janik:1999fk}
%\bibitem{cb_jp}
R.~A.~Janik and R.~Peschanski,
%``Conformal invariance and {QCD} pomeron vertices in the 1/N(c) limit,''
Nucl.\ Phys.\ B {\bf 549} (1999) 280;
%[arXiv:hep-ph/9901426].
%%CITATION = HEP-PH 9901426;%%
%
%%\cite{Bartels:2004ef}
%\bibitem{3p_blv}
J.~Bartels, L.~N.~Lipatov and G.~P.~Vacca,
%``Interactions of Reggeized gluons in the Moebius representation,''
Nucl.\ Phys.\ B {\bf 706} (2005) 391.
%%[arXiv:hep-ph/0404110].
%%CITATION = HEP-PH 0404110;%%



%gravity

%\cite{Janik:1999zk}
\bibitem{grav1}
  R.~A.~Janik and R.~Peschanski,
  %``High energy scattering and the AdS/CFT correspondence,''
  Nucl.\ Phys.\ B {\bf 565} (2000) 193;
  %[arXiv:hep-th/9907177].
  %%CITATION = HEP-TH 9907177;%%
%
%
%\cite{Polchinski:2001tt}
%\bibitem{Polchinski:2001tt}
  J.~Polchinski and M.~J.~Strassler,
  %``Hard scattering and gauge/string duality,''
  Phys.\ Rev.\ Lett.\  {\bf 88} (2002) 031601;
  %[arXiv:hep-th/0109174].
  %%CITATION = HEP-TH 0109174;%%
%
%\cite{Polchinski:2002jw}
%\bibitem{Polchinski:2002jw}
  J.~Polchinski and M.~J.~Strassler,
  %``Deep inelastic scattering and gauge/string duality,''
  JHEP {\bf 0305} (2003) 012;
  %[arXiv:hep-th/0209211].
  %%CITATION = HEP-TH 0209211;%%
%
%
%\cite{Danilov:2006fv}
%\bibitem{Danilov:2006fv}
  G.~S.~Danilov and L.~N.~Lipatov,
  %``BFKL pomeron in string models,''
  arXiv:hep-ph/0603073;
  %%CITATION = HEP-PH 0603073;%%
%
%\cite{Brower:2006ea}
%\bibitem{Brower:2006ea}
  R.~C.~Brower, J.~Polchinski, M.~J.~Strassler and C.~I.~Tan,
  %``The pomeron and gauge / string duality,''
  arXiv:hep-th/0603115.
  %%CITATION = HEP-TH 0603115;%%

%%%%%%%%%%% heavy ion 

%\cite{Policastro:2001yc}
\bibitem{ion_gravity}
  G.~Policastro, D.~T.~Son and A.~O.~Starinets,
  %``The shear viscosity of strongly coupled N = 4 supersymmetric Yang-Mills
  %plasma,''
  Phys.\ Rev.\ Lett.\  {\bf 87} (2001) 081601;
  %[arXiv:hep-th/0104066].
  %%CITATION = HEP-TH 0104066;%%
%
%\cite{Kovtun:2004de}
%\bibitem{Kovtun:2004de}
  P.~Kovtun, D.~T.~Son and A.~O.~Starinets,
  %``Viscosity in strongly interacting quantum field theories from black hole
  %physics,''
  Phys.\ Rev.\ Lett.\  {\bf 94} (2005) 111601;
  %[arXiv:hep-th/0405231].
  %%CITATION = HEP-TH 0405231;%%
%
%
%\cite{Shuryak:2005ia}
%\bibitem{Shuryak:2005ia}
  E.~Shuryak, S.~J.~Sin and I.~Zahed,
  %``A gravity dual of RHIC collisions,''
  arXiv:hep-th/0511199.
  %%CITATION = HEP-TH 0511199;%%
%

%\cite{Janik:2005zt}
\bibitem{ion_jp}
  R.~A.~Janik and R.~Peschanski,
  %``Asymptotic perfect fluid dynamics as a consequence of AdS/CFT,''
  Phys.\ Rev.\ D {\bf 73} (2006) 045013.
  %[arXiv:hep-th/0512162].
  %%CITATION = HEP-TH 0512162;%%



%%%%%%%%%%%%%%%%%%%%%%%%%%


\end{thebibliography}
\end{document}